\documentclass[12pt]{iopart}
\expandafter\let\csname equation*\endcsname\relax
\expandafter\let\csname endequation*\endcsname\relax
\usepackage{iopams}
\usepackage{bm,epsfig}
\usepackage{mathrsfs}
\usepackage{amsmath}
\include{graphics}
\usepackage{graphicx}
\usepackage{color}
\newcommand{\B}[1]{{\bm{#1}}}
\newcommand{\C}[1]{{\mathcal{#1}}}
\newcommand{\pa}{\partial}

\pdfoutput=1

\begin{document}

\pdfoutput=1

 \expandafter\let\csname equation*\endcsname\relax
 \expandafter\let\csname endequation*\endcsname\relax

\title{The Dynamics of Rapid Fracture: Instabilities, Nonlinearities and Length Scales}

\author{Eran Bouchbinder$^1$, Tamar Goldman$^2$ and Jay Fineberg$^2$}

\address{$^1$Chemical Physics Department, Weizmann Institute of Science, Rehovot 76100, Israel\\
$^2$The Racah Institute of Physics, The Hebrew University of Jerusalem, Jerusalem 91904, Israel}
\ead{jay@mail.huji.ac.il}
\begin{abstract}
The failure of materials and interfaces is mediated by cracks, nearly singular dissipative structures that propagate at velocities approaching the speed of sound. Crack initiation and subsequent propagation -- the dynamic process of fracture -- couples a wide range of time and length scales.  Crack dynamics challenge our understanding of the fundamental physics processes that take place in the extreme conditions within  the nearly singular region where material failure occurs. Here, we first briefly review the classic approach to dynamic fracture, ``Linear Elastic Fracture Mechanics'' (LEFM), and discuss its successes and limitations. We show how, on the one hand, recent experiments performed on straight cracks propagating in soft brittle materials have quantitatively confirmed the predictions of this theory to an unprecedented degree. On the other hand, these experiments show how LEFM breaks down as the singular region at the tip of a crack is approached.  This breakdown naturally leads to a new theoretical framework coined ``Weakly Nonlinear Fracture Mechanics", where weak elastic nonlinearities are incorporated. The stronger singularity predicted by this theory gives rise to a new and intrinsic length scale, $\ell_{nl}$. These predictions are verified in detail through direct measurements. We then theoretically and experimentally review how the emergence of $\ell_{nl}$ is linked to a new equation for crack motion, which predicts the existence of a high-speed oscillatory crack instability whose wave-length is determined by $\ell_{nl}$. We conclude by delineating outstanding challenges in the field.
\end{abstract}
%Uncomment for PACS numbers title message
%\pacs{00.00, 20.00, 42.10}
% Keywords required only for MST, PB, PMB, PM, JOA, JOB?
%\vspace{2pc}
%\noindent{\it Keywords}: Article preparation, IOP journals
% Uncomment for Submitted to journal title message
%\submitto{\JPA}
% Comment out if separate title page not required
\maketitle

\section{Introduction}

Our fundamental understanding of why and how materials break is, surprisingly, only about 100 years old. Estimates of the theoretical strength of materials range from $E/\pi$ to $E/8$ \cite{Mecholsky.2006}, where $E$ is the Young's modulus of the material. These estimates are simply related to the work that one would have to expend in order to overcome the potential well that holds neighboring atoms together. When pulling on a sheet of window glass, for example, one would therefore expect it to stretch at least 10\% before breaking. In practice, if you are very careful, you might be able to reach a hundredth of that strain before the glass breaks. This huge disparity between the theoretical and practical strengths of materials is entirely general in brittle materials.  Obviously there is something that these calculations are missing. The answer to this is the existence of {\em cracks} in these materials.

The modern history of fracture mechanics started with a calculation by Inglis in 1913 \cite{Inglis.13}. Inglis found that the imposition of an elliptical hole into a linear elastic sheet under uniformly applied tensile stress, entirely changed how stresses are distributed in the vicinity of the hole; externally applied stresses are amplified at the tip of the larger axis by the ratio of the large and small axes of the ellipse. This stress amplification increases without bound if the ellipse is ``squashed" to form a crack. In this limit, the stress tensor at a crack's tip, $\B \sigma$, becomes singular, increasing as $\B \sigma\! \sim\! 1/\sqrt{r}$ where $r$ is the distance from the tip. This singular behavior at a crack's tip is the basis for understanding both the strength of materials and the physics of fracture. In the example of the glass plate a crack of length $10\mu m$ is sufficient to reduce the plate's theoretical strength by a factor of $100$. Simply stated, material strength is governed by the formation and subsequent propagation of cracks. A material fails when the tip of a crack at its weakest point starts to propagate.

Models describing the {\em propagation} of cracks, or dynamic fracture, only date back to the second world war. In fact, one of the first theories of crack propagation was derived by Sir Neville Mott who was enlisted to understand the causes of brittle fracture \cite{Mott.47}. A specific problem of interest was why a significant number of rapidly constructed cargo ships, known as the Liberty ships, underwent cataclysmic failure either immediately upon or shortly after their initial launch. Since then, an immense amount of research into how and when cracks propagate has been performed. Despite this effort, there remain significant and fundamental aspects of crack propagation that we do not understand. We will show, in this review, that an improved understanding of the material behavior in the vicinity of the singular region surrounding a crack's tip has shed a new light on a number of these aspects.

\subsection{Why is fracture interesting to a physicist?}

Understanding fracture is clearly important in practical applications such as designing stronger materials and structures. The process of fracture also raises a variety of important and interesting physical questions. Crack propagation involves the integration of essential physics at an extreme range of widely varying spatial and temporal scales that link the macroscopic scales where energy is injected, stored and transported to the microscopic scales where dissipation in the form of irreversible material deformation and fracture takes place. Fracture processes therefore couple a huge range of scales.  At the smallest scales one might wish to understand how the introduction of nano-structures can affect the ultimate strength of a material. At geophysical scales one would like to understand what scales are needed to determine whether a natural fault will lose its stability and generate a massive earthquake. The study of fracture dynamics has also brought to the forefront numerous fundamental questions that pit continuum theories against discrete physics. Questions of when and how do atomic scales come into play can be important \cite{Hauch.99, Holland.99, Cramer.00}.

A crack is characterized by the singular stress fields that drive it. As nature generally will not allow ``real" singular stresses to develop, the toughness of materials is essentially determined by how these singularities are regularized. The mechanisms by which nature performs this regularization determine whether a given material can be used as a structural material (e.g. steel) or not (e.g. window glass). Furthermore, a propagating crack can rapidly reach velocities that approach material sound speeds, the speeds at which information propagates in these systems. Thus, the physics of crack propagation are closely related to questions of the formation and regularization of finite time singularities that are formed at moving fronts. These fronts are ``relativistic'' in the sense that they travel at speeds approaching the information speed in the material.

Achieving a fundamental understanding of the dynamics of fracture may also shed light on a rather broad class of conceptually related physical systems. Fracture propagation is a rather close relative of the broad class of physical problems that can be loosely characterized as ``growth" problems. These systems involve situations where a moving boundary separates two distinct phases. Examples include propagating fronts as vehicles for phase transitions (e.g. the spread of thermal convection fronts or the motion of an interface between stable and unstable regions) \cite{RBfronts,Saarloos}, Laplacian growth problems (e.g. crystal growth, Saffman-Taylor like problems of fluid invasion or imbibition) \cite{Cross_Hohenberg}, the physics of lightening \cite{lightning}, flame propagation \cite{Sivashinsky} and reaction-diffusion fronts, and the general problems of interface propagation and roughening (e.g KPZ-like problems) \cite{KPZ}. A common denominator of all of these problems involves coupling of two media, whose behavior is described by (often linear) field equations (e.g. Laplacian or diffusion equations, wave equations, Ginzburg-Landau equations) that are coupled at the boundary between the two media \cite{Cross_Hohenberg}. This boundary (or front) generally undergoes rich and varied space-time dynamics that are not a priori known, but are rather determined self-consistently from the global solution. Precisely these dynamics are what we would like to understand. In a 2D crack propagation problem, the boundary is the crack tip itself and the two crack lines left behind it. In a 3D body a crack is a 2D sheet whose leading edge is a putatively singular one-dimensional front that separates intact and fractured material. The dynamics and instabilities of this rapidly propagating singular front result from coupling the space-time behavior of the intact medium, as described by linear wave equations, to the moving boundary, which is defined by stress-free boundary conditions at the crack faces. Via the surrounding fields, the front can interact in space and time with both itself and the crack surface that it had previously formed \cite{Willis_Mochvan.95, Willis.02}. In addition, the crack front dynamics can be affected by interaction with material inhomogeneities in its path \cite{Willis_Mochvan.95,Willis.02, Morrissey.97a,Ramanathan.97,Morrissey2000,Sharon2002,bonamy.12,Bonamy.11,roux.02,Roux.03,Sharon.01} .

The study of dynamic fracture is also an interesting example of a physical system that is strongly out of equilibrium. There are few better examples for this than a nearly singular tip that is pulling a material apart at nearly the speed of information propagation. This challenging problem has forced us to extend our knowledge of materials well past the point where equilibrium properties are useful. An instructive example of this was encountered by Marder and coworkers \cite{Hauch.99,Holland.98} when they incorporated potentials that successfully describe equilibrium properties of silicon \cite{Stillinger.85} in a molecular dynamics simulation of fracture in this material. Using this potential, cracks were barely able to propagate and the energy needed to initiate fracture was 400\% larger than the measured value \cite{Hauch.99,Holland.er.98}. Later work suggested that, in possibly the most widely studied material in history, quantum mechanical calculations were needed to enable an accurate quantitative description for the {\em simplest} mode of crack propagation (the problem of a single propagating crack) \cite{bernstein.09,kermode.08}.

Modern theories of amorphous plasticity also have their roots in our collective frustration in finding a fundamental understanding of the onset and propagation of cracks \cite{Falk.11,Falk.98}. Entire classes of continuum models \cite{Langer.98,Lawn.93,98Fre,Barenblatt.59;b} that purport to describe the generic breakdown of the singular behavior at a crack's tip in amorphous materials have been shown to be unable to describe how and why a simple propagating crack becomes unstable. Failed attempts to do so have lead a number of groups to initiate theories to provide a better and fundamental description of how plastic deformation takes place in amorphous materials, where plasticity due to dislocation formation and propagation cannot play a role \cite{Falk.11,Falk.98,Langer.98, BLP07I, BLI-09, BLII-09, BLIII-09, Rycroft.2012}.

\subsection{The breakdown of linear elasticity near crack tips}

Naively, one would think that the existence of a singular stress at a crack's tip should be sufficient to cause the material at the tip to fracture. Assuming that Hooke's law describes the elastic behavior of a material (i.e. that materials are linearly elastic -- stresses are proportional to strains), singular stresses should lead to singular strains, hence to bonds at the tip of a crack being pulled apart as a crack continually extends itself. In practice, there is a threshold for crack propagation, which in essence defines the toughness of a given material. Where does this threshold come from?

The answers to this question are related to the ways that nature manages to regularize this stress singularity. Obviously, stresses and/or strains cannot really be mathematically singular so something has to happen as one approaches the small length scales that are near a crack's tip. The region where the $1/\sqrt{r}$ singularity of the stress field breaks down has been named the ``process zone''. Within the process zone all dissipative and/or nonlinear processes that take place when materials are stretched beyond the point where linear elasticity breaks down are assumed to take place. The process zone is, essentially, the rug under-which we sweep all of the ``dirty" processes for which we either lack fundamental understanding or where continuum theories break down. This nebulous region encompasses processes such as plastic (irreversible) deformations, nonlinear elastic effects, damage accumulation, visco-elastic processes, and dissipative effects due to discreteness at the atomic scale.

What is the size of the process zone? There are few direct measurements of this region and its size estimates vary over a wide range from material to material. Rough estimates are sometimes obtained by equating the singular field to a material's yield stress. These estimates are often very crude and can vary considerably. In some materials the natural cutoff for the singularity is the size of the discrete atomic scale where the continuum theory that gave rise to the stress singularity must certainly break down. In brittle materials such as crystalline silicon and even amorphous glasses $nm$ scales have been quoted for this region \cite{Lawn.93}. In brittle acrylics estimates range from 1-10$\mu m$. In very tough materials such as aluminum the process zone can reach $mm$ scales \cite{Lawn.93}, as the size of the dissipative zone is determined by the scale at which dislocations moving out of the crack tip pile up and lose their mobility.

A key assumption of fracture mechanics that provides a way to circumvent the myriad material-dependent dissipative mechanisms and nonlinear processes that are purported to take place around a crack's tip is the assumption of ``small-scale yielding". Small-scale yielding assumes that the details of the structure of the process zone can be largely ignored, if the region where these processes take place is sufficiently small compared to other characteristic length scales in a given problem. In such a case it is possible to justify a separation of scales; stress fields outside of the process zone can be described to high accuracy by the singular contribution to the linear elastic stress field, $\B \sigma\!\sim\!K/\sqrt{r}$ which, at intermediate scales, dominates all non-singular linear elastic contributions. This separation of scales enables one of the triumphs of fracture mechanics: a universal description of the functional form of the linear elastic stresses and strains surrounding the tip of a crack. As energy dissipation is confined to a small region within this universal field, predictions of material failure and the consequent motion of a crack are entirely determined by the value of $K$, which is called the ``stress intensity factor". Knowledge of $K$ is equivalent to knowing the energy flowing into to the crack's tip.  The assumption of small-scale yielding suggests an explanation for the universality of fracture behavior;  e.g. why does a brittle plastic fracture in the same way as the glass in your window -- despite that fact that the dissipative processes in both materials are vastly different.  This implied universality, which is sometimes termed ``K-dominance", suggests that all one needs to know is how to either measure or compute $K$ for a given external application of stresses in order to predict how a crack in a given material will behave.

 \subsection{Linear Elastic Fracture Mechanics (LEFM)}

Linear Elastic Fracture Mechanics, or LEFM, provides the basis for our current understanding of fracture \cite{Lawn.93, 98Fre,  Irwin.57, Rice.68, Willis.67,  Broberg.99, Slepyan.02}. In this section we will both describe LEFM and set up the notation that will enable us to later describe fracture mechanics when the assumptions of linear elasticity are relaxed.

We start by defining the motion $\B \phi(\B x,t)$, which is assumed to be a continuous, differentiable and invertible mapping between a reference (undeformed) configuration described by coordinates $\B x$ and a deformed configuration described by coordinates $\B x'$, both in ${\C R^3}$,
\begin{equation}
\label{phi}
\B x' = \B \phi(\B x,t) = \B x + \B u(\B x, t) \ ,
\end{equation}
where $\B u(\B x, t)$ is the displacement field. The components of the displacement gradient tensor $\B H$ are defined as
\begin{equation}
\label{disp_grad}
H_{ij} = \pa_j u_i \ .
\end{equation}
Note that partial spatial derivatives are assumed to be taken with respect to the reference configuration $\B x$, unless otherwise stated. However, the distinction between the reference and deformed configurations manifests itself only to nonlinear orders in $\B H$ and hence it makes no difference in the framework of linear elasticity. Within this framework one assumes $|\B H|\!\ll\!1$ and defines the infinitesimal strain tensor as
\begin{equation}
\label{infinitesimal_strain}
\B \varepsilon \equiv \frac{1}{2}(\B H + \B H^T) \ ,
\end{equation}
where the superscript `$T$' stands for the transpose of a tensor. The linear elastic energy density functional $U(\B\varepsilon)$ of isotropic materials can be expressed as
\begin{equation}
\label{U_varepsilon}
\hspace{-2cm} U(\B \varepsilon) = \frac{1}{2} \lambda \left(tr\B\varepsilon\right)^2 + \mu ~tr\B\varepsilon^2 \ ,
\end{equation}
where $\lambda$ and $\mu$ are the Lam\'e constants \cite{Landau.86}. The Cauchy stress tensor $\B\sigma$ is thermodynamically work-conjugate to $\B\varepsilon$, $\B\sigma\!=\!\pa_{\B\varepsilon} U(\B\varepsilon)\!=\!\lambda \left(tr\B \varepsilon\right) \B I \!+\! 2 \mu \B \varepsilon$ ($\B I$ is the identity tensor), which is nothing but Hooke's law \cite{Landau.86}. Substituting the latter in the linear momentum balance
\begin{equation}
\label{EOM}
\nabla \cdot \B \sigma = \rho_0 \pa_{tt}{\B u} \ ,
\end{equation}
where $\rho_0$ is the reference mass density, we obtain the standard Lam\'e equation
\begin{equation}
\mu\nabla^2{\B u^{(1)}}+(\lambda + \mu) \nabla(\nabla\cdot{\B u^{(1)}})=\rho_0\pa_{tt}{\B u}^{(1)} \ ,
\label{Lame}
\end{equation}
which is the basic equation of linear elasticity. The superscript $(1)$ was introduced to stress the fact that this equation is a first (linear) order approximation in the magnitude of $\B H$. This will be important later when higher order contributions will be discussed. Note that angular momentum balance is automatically satisfied due to the symmetry of the Cauchy stress, $\B\sigma\!=\!\B\sigma^T$ \cite{Landau.86}.

Consider now a long straight crack propagating in a 2D body and define a fixed Cartesian coordinate system $(x,y)$ such that the crack's tip propagates steadily at a velocity $v$ in the positive $x$-direction. The $y$-direction is perpendicular to the crack's faces. A crack is defined physically as composed of surfaces that cannot support stresses, i.e. by the following boundary conditions
\begin{eqnarray}
\label{BC_Cauchy}
\sigma_{xy}(r,\varphi\!=\!\pm\pi)=\sigma_{yy}(r,\varphi\!=\!\pm\pi)= 0 \ .
\end{eqnarray}
$(r,\varphi)$ is a polar coordinate system that moves with the crack tip, which is related to the rest frame by
$r\!=\!\sqrt{(x-vt)^2+y^2}$ and $\varphi\!=\!\tan^{-1}[y/(x-vt)]$. Mathematically speaking, a crack can be regarded as a moving branch cut. Under steady-state propagation
conditions we expect all of the fields to depend on $x$ and $t$ through the combination $x\!-\!vt$ and therefore $\pa_t\!=\!-v\pa_x$. $\varphi\!=\!0$ coincides with the positive $x$-direction and $\varphi\!=\!\pm\pi$ define the two opposite crack's faces. Using Hooke's law, equations (\ref{BC_Cauchy}) can be rewritten as
\begin{eqnarray}
&&\label{BC1st} -\mu\,r^{-1} \pa_\varphi u_x^{(1)}-\mu\pa_r u_y^{(1)}=0\ , \nonumber\\
&&-(\lambda+ 2\mu) r^{-1}\pa_\varphi u_y^{(1)}-\lambda \,\pa_r u_x^{(1)}=0 \ ,
\end{eqnarray}
for $\varphi=\pm\pi$.

For mode I (tensile) symmetry, i.e. $u_x^{(1)}(x,-y) \!=\! u_x^{(1)}(x,y)$ and $u_y^{(1)}(x,-y) \!=\! -u_y^{(1)}(x,y)$, in either plane-strain or plane-stress conditions \cite{98Fre,Broberg.99}, the two-term near tip asymptotic solution of equation (\ref{Lame}) with the boundary conditions of equations (\ref{BC1st}), is given as \cite{98Fre,Broberg.99}
\begin{eqnarray}
\label{u_1st_I}
\hspace{-2.3cm} u_x^{(1)}&=&\frac{2K_I}{\mu \sqrt{2\pi}D(v)}\left[(1+\alpha_s^2) r_d^{1/2}\cos{\left(\frac{\varphi_d}{2}\right)}-
2\alpha_d \alpha_s r_s^{1/2}\cos{\left(\frac{\varphi_s}{2}\right)}\right]+\frac{(\lambda+2\mu)~T~r\cos\varphi}{4\mu(\lambda+\mu)},\nonumber\\
\hspace{-2.3cm} u_y^{(1)}&=&-\frac{2K_I\alpha_d}{\mu \sqrt{2\pi}D(v)}\left[(1+\alpha_s^2 )r_d^{1/2}\sin{\left(\frac{\varphi_d}{2}\right)}-
2r_s^{1/2}\sin{\left(\frac{\varphi_s}{2}\right)} \right]-\frac{\lambda ~T~r\sin\varphi}{4\mu(\lambda+\mu)} \ .
\end{eqnarray}
Here $\alpha^2_{d,s}\!\equiv\! 1-v^2/c_{d,s}^2$, $\tan{\varphi_{d,s}}\!=\!\alpha_{d,s}\tan{\varphi}$, $r_{d,s}\!=\!r\sqrt{1-(v\sin\varphi/c_{d,s})^2}$.  $c_d\!=\!\sqrt{(\lambda+2\mu)/\rho_0}$ and $c_s\!=\!\sqrt{\mu/\rho_0}$ are the dilatational and shear wave speeds, respectively. Finally, $K_I$ is the mode I ``stress intensity factor'', $T$ is the ``T-stress'' (the amplitude of the sub-leading term in the linear elastic asymptotic expansion) and $D(v)\!=\!4\alpha_s\alpha_d-(1+\alpha_s^2)^2$. The latter vanishes at the Rayleigh wave-speed $c_R$, $D(v\!=\!c_R)\!=\!0$. Note that the term proportional to $K_I$ in equations (\ref{u_1st_I}) gives rise to the famous $1/\sqrt{r}$ displacement-gradients (and stress) singularity discussed above \cite{98Fre, Broberg.99}. Similar expressions are obtained under global shear loading (mode II fracture), where the mode II stress intensity factor $K_{II}$ appears instead of $K_I$. The role of $K_{II}$ in determining the direction of crack propagation will be discussed later.

One immediate implication of the solution in equations (\ref{u_1st_I}) is that the crack that is represented by a straight branch cut in the reference configuration becomes parabolic near its tip in the deformed configuration
\begin{eqnarray}
\label{parabola}
\phi_x(r,\pm\pi) = - \chi_1 ~\phi^2_y(r,\pm\pi)\ ,
\end{eqnarray}
obtained by expressing $\phi_x(r,\pm\pi) \!=\! -r + u_x(r,\pm\pi)$ as a function of $\phi_y(r,\pm\pi) \!=\! u_y(r,\pm\pi)$ (see equation (\ref{phi}) for the definition of $\B \phi$), where the curvature $\chi_1$ can be easily read off equations (\ref{u_1st_I}) in terms of $K_I$ and $T$. This is the so-called parabolic ``crack tip opening displacement'' (CTOD). It is important to note that the values of $K_I$ and $T$ in equations (\ref{u_1st_I})-(\ref{parabola}) depend on the driving of the system, specifically the globally applied stresses. These, of course, cannot be determined by asymptotic analysis. In the next subsection we will see how the stress intensity factor is related to both the applied stresses and the energy flux into the crack tip, and consequently to the crack's motion.

\subsection{Fracture initiation and energy balance}

When a crack extends, it forms two new surfaces at an energy cost. We define the fracture energy, $\Gamma$, as the energy needed per unit extension of a crack to create these surfaces. Griffith suggested ``energy balance" as a criterion for crack stability \cite{Griffith.20}. The resulting ``Griffith criterion" states that a crack will lose stability when the change in potential energy $U$ in the surrounding medium (including the loading machine) released by the crack upon an infinitesimal extension, $\delta l$, of its length $l$ surpasses $\Gamma$. While the ``Griffith condition'', $\pa U/\pa l\!>\!\Gamma$, avoids the need to explicitly take into account the nature of the stresses at a crack's tip, it is clear that the energy needed to propagate a crack and the form of the stress field surrounding a propagating crack must be related. (We note in passing that Griffith considered only the bare surface energy $2\gamma$ as the cost of crack initiation. The fracture energy $\Gamma$ is a generalization of the concept of surface energy, see below for an additional discussion of this point). As we have seen in equation (\ref{parabola}), sufficiently close to the crack's tip, the only part of the asymptotic solution that can contain information about the global loading conditions is the stress intensity factor, $K_I$. Considering a nearly static crack, Irwin indeed showed that the energy flowing into the crack's tip per unit crack extension, $G(v)$, is given (under plain-strain conditions \cite{Landau.86}) by \cite{Irwin.57}
 \begin{equation}
G _{v \rightarrow 0}=  \frac{1-\nu ^2}{E}K_I^2 \ .
\label{G_stat}
\end{equation}
In the presence of shear stresses near the crack tip an additional contribution proportional to $K^2_{II}$ appears \cite{98Fre}.

The quantity $G$ in equation (\ref{G_stat}) is called the {\em energy release rate} (even though it involves no rate, its dimensions are energy per unit area). It quantifies the amount of energy per unit fracture surface that is flowing into the tip of a crack, where it is dissipated.
The generalized Griffith condition for fracture initiation therefore reads
\begin{equation}
G _{v \rightarrow 0}=  \frac{1-\nu ^2}{E}K_I^2 =\Gamma_{v \rightarrow 0}.
\label{initiation}
\end{equation}
Equation (\ref{initiation}) demonstrates that the generalized Griffith condition for the stability of a static crack is equivalent to positing a critical value of the stress intensity factor.
The dimensions of $K_I$, as can be seen in equation (\ref{u_1st_I}), are those of $\hbox{stress}\!\times\!\sqrt{\hbox{length}}$.
The explicit value of $K_I$ can be calculated for a static crack in a given loading configuration. For example, when a constant tensile stress $\sigma_\infty$ is applied to the remote boundaries of a
 large sheet in which a crack of length $l$ exists, $K_I \!\propto\! \sigma_\infty \sqrt{l}$. If the same uniform stress is applied to an infinitely long strip of width $2b$ then $K_I \!\propto\! \sigma_\infty \sqrt{b}$.

Whereas both $K_I$ and $G$ are loading-dependent quantities, $\Gamma$ is considered to be a material-dependent quantity that simply specifies a given material's resistance to being broken. $\Gamma(v)$ can in fact be a rate-dependent function and is, therefore, generally dependent on the instantaneous crack velocity $v$. $\Gamma(v)$ encompasses all of the dissipative processes that take place within the process zone. It is {\em not} simply the energy cost of breaking a single plane of material bonds, i.e. the surface energy $2\gamma$, as a variety of possible dissipative material-dependent processes need to take place to enable a bond to be finally broken. Often these processes require orders of magnitude more energy that the simple fracture of the bonds that hold a material together. For example, the fracture energy of a brittle acrylic such as PMMA (poly-methyl-methacrylate or ``plexiglas") requires $500-1000 ~\hbox{J/m}^2$ whereas breaking a plane of dense carbon bonds would typically require about two orders of magnitude less energy. The huge amount of  ``extra" energy that goes into $\Gamma$ for PMMA is thought to result from  plastic deformation of the polymer that is a necessary condition for the separation of the polymer chains that compose PMMA.

\subsection{Equations for crack growth rate}

Equation (\ref{initiation}) tells us what is needed to initiate fracture. What happens once a crack starts to propagate? The condition of dynamic energy balance, $G(v)\!=\!\Gamma(v)$, which further extends the generalized Griffith criterion to all velocities (crack growth rates $v$) is the basis for the formulation of a dynamic theory of fracture \cite{98Fre}. Once $G(v)$ is calculated and $\Gamma(v)$ is either calculated or measured, one is able to predict $v$ as a function of variables such as the crack length and parameters such as any externally applied stresses {\em as long as the crack path is known a priori}. That means that the scalar equation $G(v)\!=\!\Gamma(v)$ is, in principle, capable of determining the crack growth rate $v$, but not the crack's direction of propagation. The analytic calculation of $G(v)$ is generally very difficult. Two crack configurations for which it has been performed are:
\begin{itemize}
\item A medium of effectively infinite spatial extent for any combination of forces applied to the crack faces (or any problem that can be mapped to such a problem) \cite{98Fre}.
\item A semi-infinite crack propagating at any velocity, $v$, within an infinitely long strip whose finite boundaries are subjected to constant displacement conditions \cite{Marder.91.prl}.
\end{itemize}

Both of the above calculations are limited to simple, perfectly straight cracks that have not undergone any path instabilities, such as oscillations or branching. Propagating cracks in each of the above calculations have the same form of asymptotic stress tensor field, $\B\sigma \!\propto\! K_I(v)/\sqrt{r}$, but with different dependencies of $K_I(v)$ in terms of the system parameters and geometry. Below we will briefly review the main features of each of the resulting equations of crack growth rate.

\subsubsection{The motion of a crack in an infinite medium}

The first calculation is due to Eshelby \cite{Eshelby.69}, Freund \cite{98Fre}, Kostrov \cite{Kostrov.66,Kostrov.74}, and Willis \cite{Willis.90}. It strictly applies to a
perfectly straight semi-infinite crack in an infinite plate propagating at a steady or non-steady rate $v$, with loads applied to the crack faces. The latter can can be mapped by superposition to a broader class of loading configurations, taking advantage of the linearity of the field equations and boundary conditions in (\ref{Lame}) and (\ref{BC1st}). This calculation is also valid for bodies of finite size for times that are shorter than the time needed for waves to bounce off of the boundaries and return to interact with the crack. With these restrictions the calculation is exact, and holds with remarkable generality. It is commonly thought that this is the {\em only} equation of motion for a dynamic crack, but this is not the case (see below).

The full details of the derivation can be found in Freund's book \cite{98Fre} and with a less detailed review in \cite{Fineberg.99}. The asymptotic displacement fields $\B u$ for a crack propagating in an infinite plate have previously been presented in equation (\ref{u_1st_I}), as a function of the stress intensity factor $K_I(v)$. The dynamic theory predicts that the dynamic stress intensity factor $K_I(v)$ can be written as
\begin{eqnarray}
K_I(v)=k(v) K_s(\hbox{loading},l)\ ,
\label{K(v)}
\end{eqnarray}
with
\begin{eqnarray}
k(v)\simeq (1-v/c_R)/\sqrt{1-v/c_d} \ .
\label{k(v)}
\end{eqnarray}
$k(v)$ is a universal function of the crack growth rate $v\!=\!\dot{l}$ alone.  $K_s(\hbox{loading},l)$ depends on the loading configuration and the crack length $l$, but not on $v$ \cite{98Fre}. The computation of the latter might be difficult and may require numerical techniques, but it involves no dynamics. Equation (\ref{K(v)}) is remarkable because it decomposes a dynamic quantity $K_I(v)$ into a purely dynamic universal function $k(v)$ and a problem-specific quantity $K_s(\hbox{loading},l)$ that does not involve dynamics at all (e.g under applied tensile stress far from the crack faces $\sigma_\infty$, $K_s\!\sim\!\sigma_\infty\sqrt{l}$). Thus, a dynamic problem is reduced to a non-dynamic one.

For a crack moving at any velocity $v$ in an infinite plate, the energy balance of equation (\ref{initiation}) generalizes to
\begin{equation}
\Gamma(v) = G(v)=  \frac{1-\nu ^2}{E}K^2_I(v) A(v)= \frac{(1-\nu ^2)K^2_s(l)}{E}  k^2(v)  A(v) \label{G(v)} \ ,
\end{equation}
where $A(v)$ is yet another known universal function \cite{98Fre} and for simplicity we suppressed the loading dependence of $K_s$. To a good approximation we have $k^2(v) A(v) \!\simeq\! 1-v/c_R$, which yields \cite{98Fre}
\begin{eqnarray}
\Gamma(v) \simeq \frac{1-\nu ^2}{E}K_s^2(l) (1-v/c_R)~~ \Rightarrow ~~ v \simeq c_R \left[ 1-\frac{\Gamma(v) E}{(1-\nu^2)K_s^2(l)}\right] \ .
\label{eqn_motion}
\end{eqnarray}
The crack growth equation in (\ref{eqn_motion}), which is an equation for the time evolution of the crack tip location $l(t)$, has the following rather interesting features:
\begin{itemize}
  \item The quantity $G_s(l)\!\equiv\!(1-\nu^2)K_s^2(l)/E$ can be interpreted as an effective thermodynamic force for crack motion. No motion takes place if $G_s\!<\! \Gamma$. For $\Gamma \!\rightarrow\! 0$ (no resistance to crack propagation) or $G_s \!\rightarrow\!\infty$ (infinite effective driving force), a crack will accelerate to a {\em finite} limiting speed equal to the Rayleigh wave speed $c_R$.
  \item Equation (\ref{eqn_motion}) depends on the instantaneous tip location $l$ and its instantaneous speed $v\!=\!\dot{l}$, but not on higher order time derivatives. In particular, the absence of an inertia-like acceleration term $\dot{v}\!=\!\ddot{l}$ in this equation implies that the crack tip can be treated as a massless particle/defect that responds {\em instantaneously} to any change in either $\Gamma$ or the driving force $G_s$. From this perspective, equation (\ref{eqn_motion}) takes the form ``$v\!=\!F/\eta$'', where the RHS is the ratio of the net driving force ``$F$'' over an effective mobility ``$\eta$''. A term ``$m \dot{v}$'', where ``$m$'' is an effective mass, is missing from this equation.
  \item Comparison of equations (\ref{G_stat}) and (\ref{eqn_motion}) reveals that the dynamic contributions to the energy release rate serve to reduce its value relative to its quasi-static limit, $G(v\!=\!0)$. This is basically due to the increase in kinetic energy required to move the surrounding material away from the crack faces as the crack tip progresses. This energy cost diverges as $v \!\rightarrow\! c_R$ and is the heart of the existence of a finite limiting velocity for a crack.
\end{itemize}

\subsubsection{The motion of a crack in an infinitely long strip}

When a crack is propagating within an infinitely long strip, its motion is described by a qualitatively different equation derived by
Marder \cite{Marder.91.prl}. This crack configuration is commonly used to obtain model-independent measurements of $\Gamma(v)$ \cite{Sharon.2.96,Sharon1996,Baumberger.06}. The rationale behind this is as follows. Consider a semi-infinite crack moving at a constant steady-state velocity $v$ in an infinitely long strip of width $2b$ (in the $y$ direction) along its symmetry axis (in the $x$ direction). Energy is stored in the strip by displacing its boundaries at $y\!=\!\pm b$ by a constant amount. Far ahead of the crack tip ($x \!\rightarrow\! +\infty$) the strip stores a constant energy per unit length, $W$, whereas far behind the crack tip ($x \!\rightarrow\! -\infty$) all of this energy has been transformed into creating new crack surfaces. Under these conditions energy balance requires that $\Gamma(v)\!=\!W$ since translational invariance simply implies that when a unit length of new surface is formed at the crack tip an amount $W$ of energy is released at $x\!\rightarrow\!+\infty$ while a new unit length of relieved stresses is added at $x\!\rightarrow\! -\infty$. Marder considered a crack's dynamics within the strip under non-steady conditions \cite{Marder.91.prl}. Performing a perturbative analysis where the
dimensionless acceleration $b\dot{v}/c_d^2$ was assumed to be small, energy balance yielded
\begin{equation}
\label{marder} G \simeq
W\left[1-\frac{b\dot{v}}{c_d^2}f(v)\right] \simeq W\left[1-\frac{b\dot{v}}{c_d^2}\left(1-\frac{v^2}{c_R^2}\right)^{-2}\right]=\Gamma(v) \ .
\end{equation}

Let us compare equations (\ref{eqn_motion}) and (\ref{marder}). One obvious difference is that the former depends on the crack's tip position $l$, while the latter is independent of it, but rather depends on the strip's half width $b$. More importantly, in contrast to equation (\ref{eqn_motion}), equation (\ref{marder}) tells us that the motion of a crack in a strip explicitly depends on its acceleration $\dot{v}$. Rewriting equation (\ref{marder}) as
\begin{equation}
\label{marderB}
\frac{f(v)b}{c_d^2}\dot{v}=1-\frac{\Gamma(v)}{W} \ ,
\end{equation}
we observe that this equation takes the form ``$m(v)\dot{v}\!=\!F$'', where the net force ``$F$'' is proportional to the difference between the effective thermodynamic force $W$ and the crack propagation resistance $\Gamma(v)$, and the velocity-dependent effective mass ``$m(v)$'' is proportional to $f(v)\!\simeq \!(1-v^2/c_R^2)^{-2}$. One family of steady state solutions is obtained when $\Gamma(v)\!=\!W$. However, as $\Gamma(v)$ is a material function that is bounded for $0\!\le\!v\!\le\!c_R$, one can externally set $W$ such that $W\!>\!\Gamma(v)$ for every $0\!\le\!v\!\le\!c_R$. In this case, the solution approaches $v\!\to\!c_R$ in such a way that the effective mass diverges, $m(v)\!\sim\!f(v)\!\to\!\infty$, and $\dot{v}\!\to\!0$ while their product remains finite.

While both equations (\ref{eqn_motion}) and (\ref{marder}) (or equivalently equation (\ref{marderB})) are manifestations of energy balance at the crack tip and both predict $v\!\to\!c_R$ for sufficiently intense loadings, the underlying physics is quite different. In an infinite system elastic waves always outrun the crack tip whereas in a strip these waves are reflected back into the system and forced to interact with the crack. This qualitative difference in crack dynamics predicted by equation (\ref{eqn_motion}) and the dynamics that occur when a crack is able to interact with its ``history" and acquire effective ``inertia" (e.g. as in equation (\ref{marderB})) is often overlooked.

\subsection{Using soft materials to test LEFM}

Whereas the LEFM predictions of fracture onset (e.g. equation (\ref{initiation})) have been experimentally validated for many materials (in fact this equation is used to {\em measure} the fracture energy $\Gamma$ at the onset of fracture), only a few direct experimental tests of the LEFM predictions for the {\em dynamic} behavior of cracks have been performed \cite{Parleton,Sharon.99}. Historically, early measurements showing the inability of the maximal observed propagation velocity of a crack in brittle materials to approach $c_R$ had suggested that something was seriously wrong with our fundamental understanding of rapid crack dynamics, but it was unclear where the root of the problem lay. Later on, the problem was attributed to the existence of instabilities, which were observed both experimentally \cite{Cramer.00, Fineberg.99, Sharon.2.96,Ravi-Chandar.84c,Fineberg.91,Gross.93,Sharon.95,Boudet.95,Livne.05, Yang.96, Ravi-Chandar.97} and numerically/theoretically \cite{Abraham.94,Xu.94,Marder.93.prl,Miller.99, Gao.93,Ching.94,Marder.95.jmps,Ching.95,Gao.96,Adda-Bedia.Ben-Amar.96,Ching.96a,Ching.96b,Ching.96c,Gumbsch.97,
Brener.98,Sander.99,Adda-Bedia.99,Boudet.00,Pla.00,Heizler.02,Buehler.03,Bouchbinder.04,Adda-Bedia.04,Bouchbinder.05a,Bouchbinder.05b,Buehler2006,Pilipenko.07,Bouchbinder.07}. The reader is referred to \cite{Fineberg.99} for a comprehensive review of some of these works.

The main experimental obstacle to direct tests of the theoretical predictions is the very high propagation velocity of cracks in typical brittle materials. In brittle polymers $c_s$ surpasses 1000 m/sec and in soda-lime  glass (window-pane glass) shear velocities surpass 3000 m/sec. At these extreme velocities, direct visualization of displacements or material deformation in the vicinity of the crack tip is extremely difficult, necessitating extremely rapid high-speed cameras with frame rates in excess of $10^6$ frames/sec. While such high speed cameras exist, measurement of near-tip crack deformations is further complicated by the microscopic size of the singular region in such materials. For example, a sheet of brittle glass typically fractures at strains of less than 0.1\% which in a sheet of 200 mm extension would result in a total crack tip opening of $\sim\!100\!-\!200 \mu$m. Moreover, the singular region near the crack tip will be of the order of 1-10 nm. Thus, real-time visualization of the singular part of the strain field surrounding the tip of a crack at reasonable spatial resolution would necessitate much better than $1\mu$m spatial resolution at frame rates exceeding $10^6$. This is prohibitively difficult with existing imaging technology.

The above obstacles can be circumvented by the use of soft materials ($E\approx 100$ kPa), such as aqueous gels. These neo-Hookean materials are both elastic and can be made to be brittle, by tuning their chemical composition \cite{Livne.05}. The shear wave speed of soft Polyacrylamide gels (13.8\% acrylamide, 2.6\% bis-acrylamide) can be tuned to be approximately 6 m/sec, nearly 3 orders of magnitude slower than ``standard" brittle materials (see Table \ref{table}). In addition, experiments using polyacrylamide gels indicate that these materials can be made to be relatively ``tougher" than other brittle materials; typically fracturing at 5-10\% strains \cite{Livne.05,Livne.2010,Goldman.2010}. Under these conditions, the near-tip singular region can exceed 1 mm in size thus enabling direct visualization by means of moderately fast video cameras.

\begin{table}
\caption{\label{table} Material properties of common brittle materials.}
\footnotesize\rm
\begin{tabular*}{\textwidth}{@{}l*{15}{@{\extracolsep{0pt plus12pt}}l}}
\br
Material&Young's modulus (MPa)&Rayleigh wave speed, $c_R$ (m/sec)\\
\mr
\verb"Soda-Lime glass"&70000&3400\\
\verb"PMMA"&4000&950-1200\\
\verb"Polyacrylamide"&0.1&5.6\\
\br
\end{tabular*}
\end{table}

\begin{figure}[ht]
\includegraphics[width=0.95\columnwidth,clip=true,keepaspectratio=true]{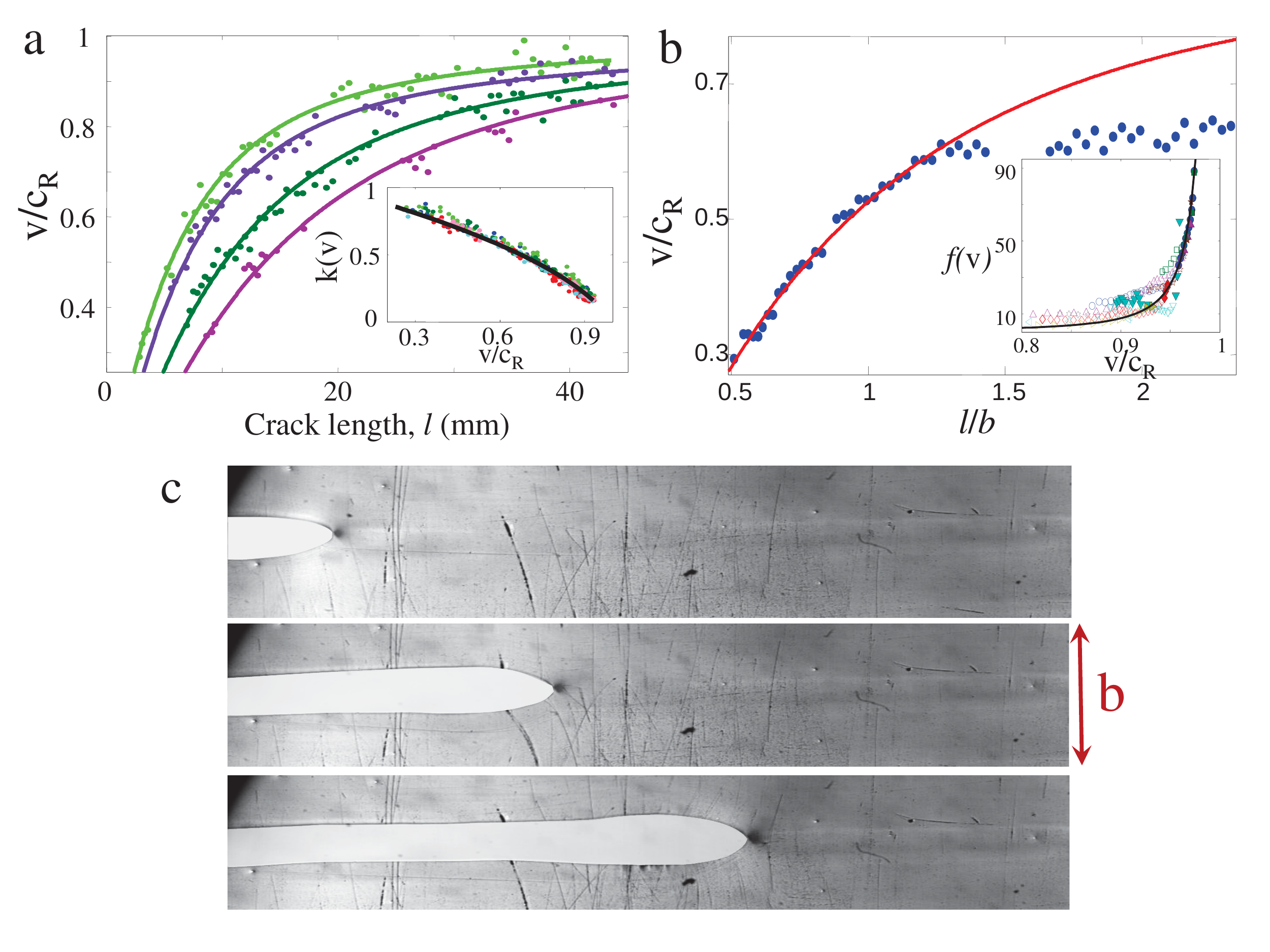}
\caption{Both equations of motion predicted by LEFM for an infinite medium (equation (\ref{eqn_motion})) and for an infinite strip (equation (\ref{marder})) are in excellent quantitative agreement with experiments performed using a polyacrylamide gel \cite{Goldman.2010}. (a) Comparison of equation (\ref{eqn_motion}) for crack dynamics in an infinite medium for different applied values of $\sigma_\infty$. The inset shows a comparison of the universal function $k(v)$ in equation (\ref{k(v)}) with experiments. (b) Crack dynamics for a finite strip compared to infinite medium prediction (solid line) as a function of the crack length, $l$, normalized by the strip width, $2b$. Once $\l\approx b$ the dynamics change dramatically and the crack dynamics are no longer consistent with equation (\ref{eqn_motion}) (red line). (inset) A comparison of the function $f(v)$ predicted in equation (\ref{marder}) \cite{Marder.91.prl} (solid line) with experimental measurements in a strip geometry. (c) a sequence of photographs of the crack tip profile during the transition from an effectively infinite medium to a strip geometry. Note how the parabolic form of the tip at $l\sim 0.6b$ (top) transitions at  $l\sim 1.4b$ (center) to a ``tadpole-like" form at $l\sim 2.2b$ (bottom) as the crack interacts with the waves reflected from the vertical boundaries. The parallel crack faces well behind the tip correspond to the displacement of the vertical boundaries. The arrow length is that of the half-width, $b$, of the strip.  See \cite{Goldman.2010} for details of the experiments.}  \label{lefm_comparison}
\end{figure}

Can these ``novel" materials indeed be used to investigate the brittle fracture of ``standard" materials? The answer is, emphatically, yes! As figure \ref{lefm_comparison} clearly demonstrates, crack dynamics are in excellent agreement with both the predictions for an infinite medium (equation (\ref{eqn_motion})) and for propagation in an infinitely long strip (equation (\ref{marder})). In the experiments on gels described in figure \ref{lefm_comparison}, crack instabilities were suppressed (see section \ref{oscillatory}) so as to enable single cracks to attain velocities in excess of $0.9c_R$. Experimental details can be found in \cite{Goldman.2010}. It is worth noting that the experiments in these comparisons were performed with {\em no} adjustable parameters. Excellent quantitative agreement was obtained for the crack dynamics in both cases as well as for the universal functions $k(v)$ and $f(v)$ that were predicted in, respectively, equations (\ref{eqn_motion}) and (\ref{marder}). Thus, not only do these types of materials correspond precisely to LEFM predictions for the motion of a single crack, these measurements have provided the most precise testing grounds over the widest range of crack velocities to date.

The excellent compliance with LEFM predictions with experiments serves to validate both equations of crack growth rate (\ref{eqn_motion}) and (\ref{marder}), and could essentially serve as a ``definition" of what we mean by a brittle material. The underlying assumptions of both equations of crack growth rate are:
\begin{itemize}
 \item Only a single ``simple" straight crack is propagating in the material.
 \item Small-scale yielding is obeyed in the material, i.e. all of the dissipation that takes place within the fracture process zone is confined within the small singular region surrounding the crack tip.
 \item The energy dissipated within the region per unit crack extension (the fracture energy $\Gamma$) is a material-dependent function of the instantaneous crack velocity $v$.
 \end{itemize}

\section{The limitations of LEFM: Crack instabilities}
\label{instabilities}

\subsection{The micro-branching instability}

In the previous section we demonstrated that, as long as fracture occurs via a single ``simple" crack, we attain an excellent picture of its overall dynamics by balancing the energy driving it, $G$, with the dissipation at its tip, $\Gamma$. Unfortunately, this is not a complete picture of brittle fracture. Experiments in a variety of different brittle materials have shown that beyond a critical velocity of about $v_c\!\approx\!0.4c_R$ a single crack becomes unstable via an instability coined the ``micro-branching" instability \cite{Fineberg.99,Ravi-Chandar.84c,Fineberg.91,Sharon.95,Gross.93,Livne.05,Hauch.98}. Beyond $v_c$ a single crack can undergo spatially local ``frustrated" crack branching events, where a single ``mother" crack gives birth to successive short-lived ``daughter" cracks. The small ``micro-branches" formed by the daughter cracks are confined to a very short range surrounding the crack tip ($1\!-\!100\mu$m in PMMA). The micro-cracks become progressively longer and more dense with increasing $G$. In fact, sufficiently beyond $v_c$  the amount of net fracture surface formed by both the mother and daughter cracks can increase extensively; a fracture surface increase of over an order of magnitude was measured in PMMA as $v$ increased from $v_c$  to $1.5v_c$ \cite{Sharon1996,Sharon.2.96}.
\begin{figure}[ht]
\includegraphics[width=0.95\columnwidth,clip=true,keepaspectratio=true]{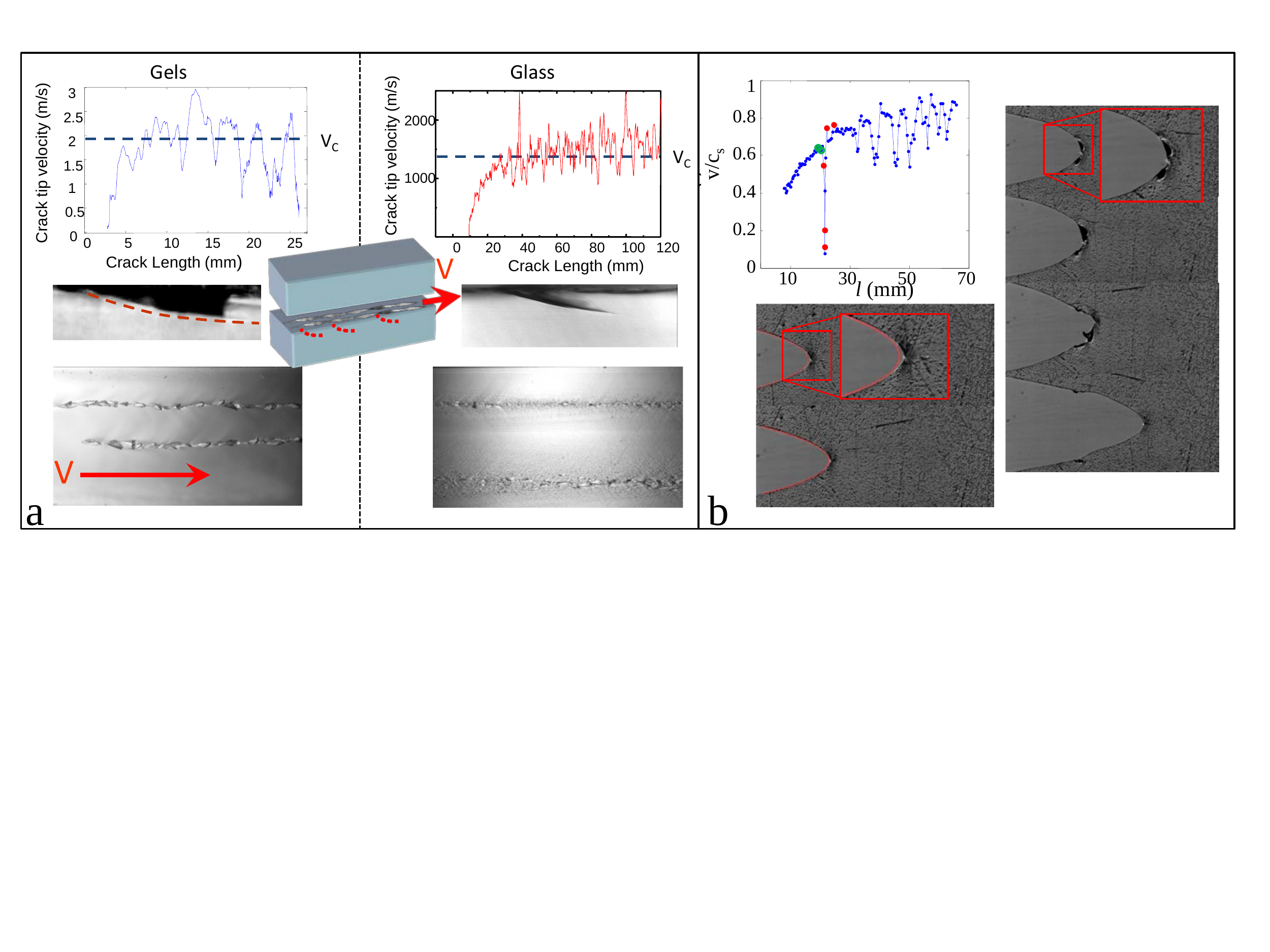}
\caption{Universality of the micro-branching instability (a) A comparison between the instability in both soft polyacrylamide gels (left) and soda-lime glass (right). Once a crack attains a critical velocity of $v\!\approx\! 0.4c_R$, a single ``simple" crack may become unstable to the micro-branching instability. At this point (top) the instantaneous velocity of the crack undergoes violent oscillations that correspond to the formation of subsurface frustrated micro-branches (center), where side views ($xy$ plane) of micro-cracks in both materials are presented ($x$ is the propagation direction, $y$ is the loading direction and $z$ is the direction along the crack front). Micro-branches are very similar in both materials, characterized by a power-law functional form. (bottom) Photographs of the resulting fracture surface ($xz$ plane) formed by cracks propagating from left to right at $v\!\sim\!0.5c_R$. The chains of structures on the fracture surface correspond to chains of micro-branches that are aligned in the propagation direction $x$ and highly localized in the $z$ direction. The vertical dimensions of the photographs are about $0.5mm$. The cartoon in the center describes the geometry; the $xz$ fracture surface is formed by a propagating cracks whereas  micro-branches (dotted lines) extend below the surface. (b) A series of photographs of the quasi-parabolic profile of running cracks in polyacrylamide gels. The two lower left panels depict the parabolic crack tip opening profile of ``simple" cracks propagating just prior to the onset of the micro-branching instability, as predicted by equation (\ref{parabola}).
The photographs correspond to the green dot on the velocity vs. crack length figure in the upper right corner. Right panels: 4 profiles of cracks undergoing the micro-branching instability, corresponding to the red dots in the velocity measurements. Note the small micro-cracks formed at the tip of the main crack.}  \label{instab_figure}
\end{figure}

A brief review of this instability is presented in figure \ref{instab_figure}, where typical examples of fracture in both soft gels and glass are described. The figure clearly demonstrates the degree of universality of this instability. Figure \ref{instab_figure}a shows that nearly every feature that characterizes the fracture process in glass is quantitatively similar (when properly normalized by the relevant velocity and length scales) to those features observed in the fracture of brittle gels \cite{Livne.05}. These features include both the functional form of the subsurface profile of the micro-branches and the corresponding structure observed on the fracture surface that is formed once the instability is excited. In both materials, micro-branches are highly localized in the $z$ direction (the depth of the sample, perpendicular to both the propagation direction $x$ and the loading direction $y$). In addition, micro-branches form directed chains along the fracture surface in the direction of propagation; each micro-branch triggering the next one along the chain. This localized structure along $z$ (crack front direction) highlights the intrinsically 3D nature of the micro-branching instability.  This may explain why descriptions of the instability using generalized energy balance arguments for branched cracks in 2D media have not yielded quantitative agreement with measurements of instability thresholds \cite{Bouchbinder.05a,Adda-Bedia.05}.

The slowing down of the fracture process that was so useful in quantitatively testing the predictions of LEFM for ``simple" cracks (e.g. figure \ref{lefm_comparison}) is also a distinct advantage in investigating the properties of this instability. A typical series of photographs of the instantaneous crack-tip profiles that bracket the onset of the micro-branching instability is presented in figure \ref{instab_figure}b. Prior to the instability onset, the crack tip propagates smoothly and is described well by the parabolic form predicted by equation \ref{parabola}. When micro-branching takes place, the overall propagation becomes jerky and the form of the tip becomes blunted, as the branches compete with the main crack, before being ``outrun" by the winner.

\subsection{Oscillatory instability in rapid fracture}\label{oscillatory}

\begin{figure}[ht]
\includegraphics[width=0.95\columnwidth,clip=true,keepaspectratio=true]{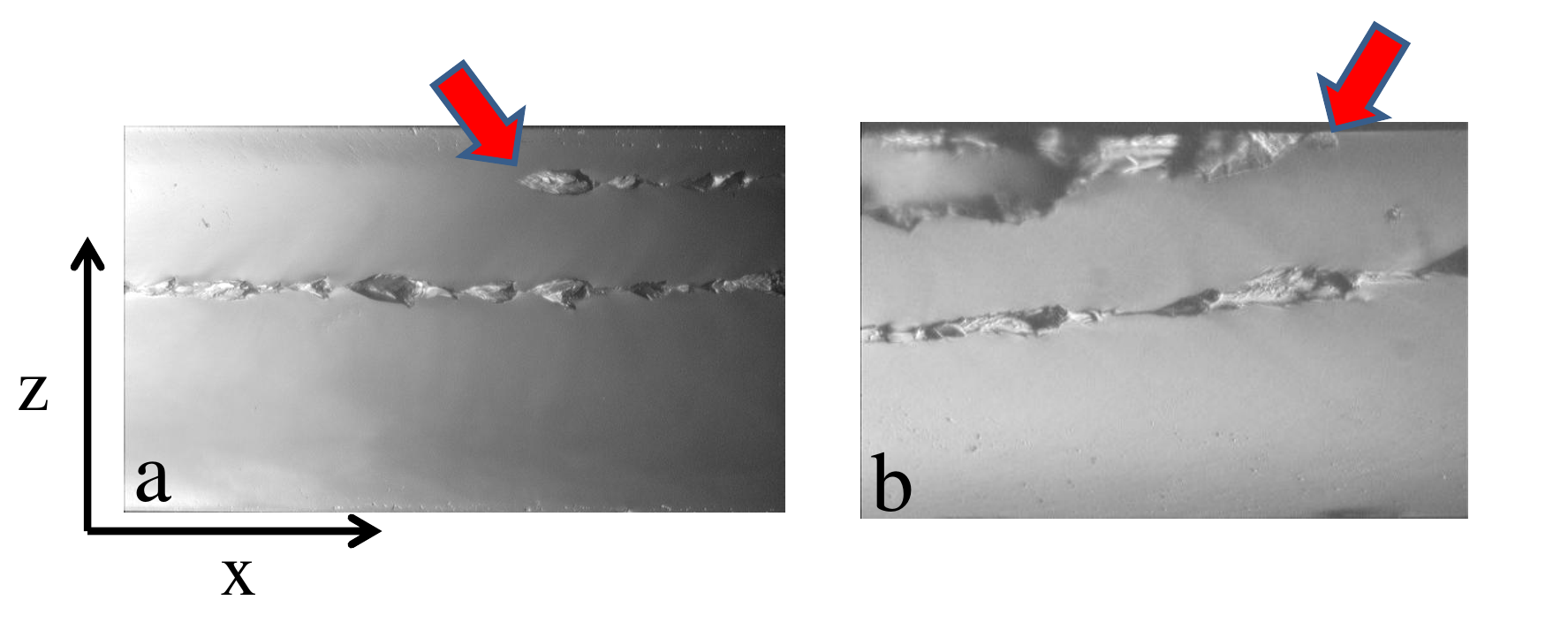}
\caption{The micro-branching instability can be suppressed by using very thin samples and high accelerations. (a) Micro-branching is an activated process. The arrow indicates a chain of micro-branches generated at a nucleation site. (b) Once a chain of micro-branches encounters the free surface of the sample (arrow), it is annihilated.
In both (a) and (b) fracture surfaces of width 0.5 mm, created by cracks propagating from left to right, are shown.}  \label{suppression}
\end{figure}

The use of gels has enabled us to both observe and manipulate cracks at a level that is unprecedented in the study of fracture in ``standard" brittle materials. For example, in gels it is possible to drive the system at very large strains, prior to fracture, and therefore achieve high acceleration rates. Applying these high accelerations, Livne et al. \cite{Livne.05} found that, while the
value of $v_c/c_R$ at the lowest acceleration rates corresponded to the value observed in PMMA and glass, $v_c$ is -- on average -- a roughly linearly increasing function of the acceleration. This same study also established that the transition to micro-branching is highly hysteretic with features that are characteristics of an activated process accompanying a first order phase transition. A typical activation of a micro-branch chain is demonstrated in figure \ref{suppression}a. Beyond a minimum value of $v_c \!\approx\! 0.4c_R$, there is a bistable region of velocities in which either a single or multi-crack state can exist. In this region, the instability may be triggered when random perturbations surpass a critical threshold for activating the first micro-branch. This yields a finite probability to bifurcate in each time interval for $v\!>\!0.4c_R$.

The sub-critical nature of the instability provides us with a means to achieve {\em single crack} states at unprecedentedly high velocities \cite{Livne.07}. The micro-branching instability is suppressed by driving cracks at very high acceleration rates. Furthermore, once the instability is activated, a chain of micro-branches that encounters a plate boundary (free surface of a plate) will disappear as shown in figure \ref{suppression}b. Thus, when an experiment is performed using very thin plates, activation centers that can trigger the instability are relatively sparse and, if the instability is triggered, the resulting chain of micro-branches quickly encounters a free surface at the edge of the fracture plane and disappears. This ``trick" was used to obtain the unprecedented range of crack velocities used to validate the LEFM equations of motion in figure \ref{lefm_comparison}.

Despite these ``tricks", a close look at figure \ref{lefm_comparison} indicates that it is still impossible to reach the asymptotic velocity of $c_R$. When the micro-branching instability is suppressed, a new and unexpected oscillatory instability is observed at a critical velocity of about $0.9c_s$ \cite{Livne.07, Goldman.12}. Similar wavy cracks traveling at velocities beyond the shear-wave speeds have also been observed in experiments on latex sheets under biaxial tension at extremely high ($>\!100\%$) strains \cite{Deegan.02.rubber,Petersan.04,Marder.JMPS.2006}. The oscillatory instability in the gels, however, may be qualitatively different in nature. These (the gel) oscillations occur at clearly subsonic velocities and are driven by purely uniaxial tension at relatively small ($\sim 15\%$) strains.
\begin{figure}[ht]
\includegraphics[width=0.95\columnwidth,clip=true,keepaspectratio=true]{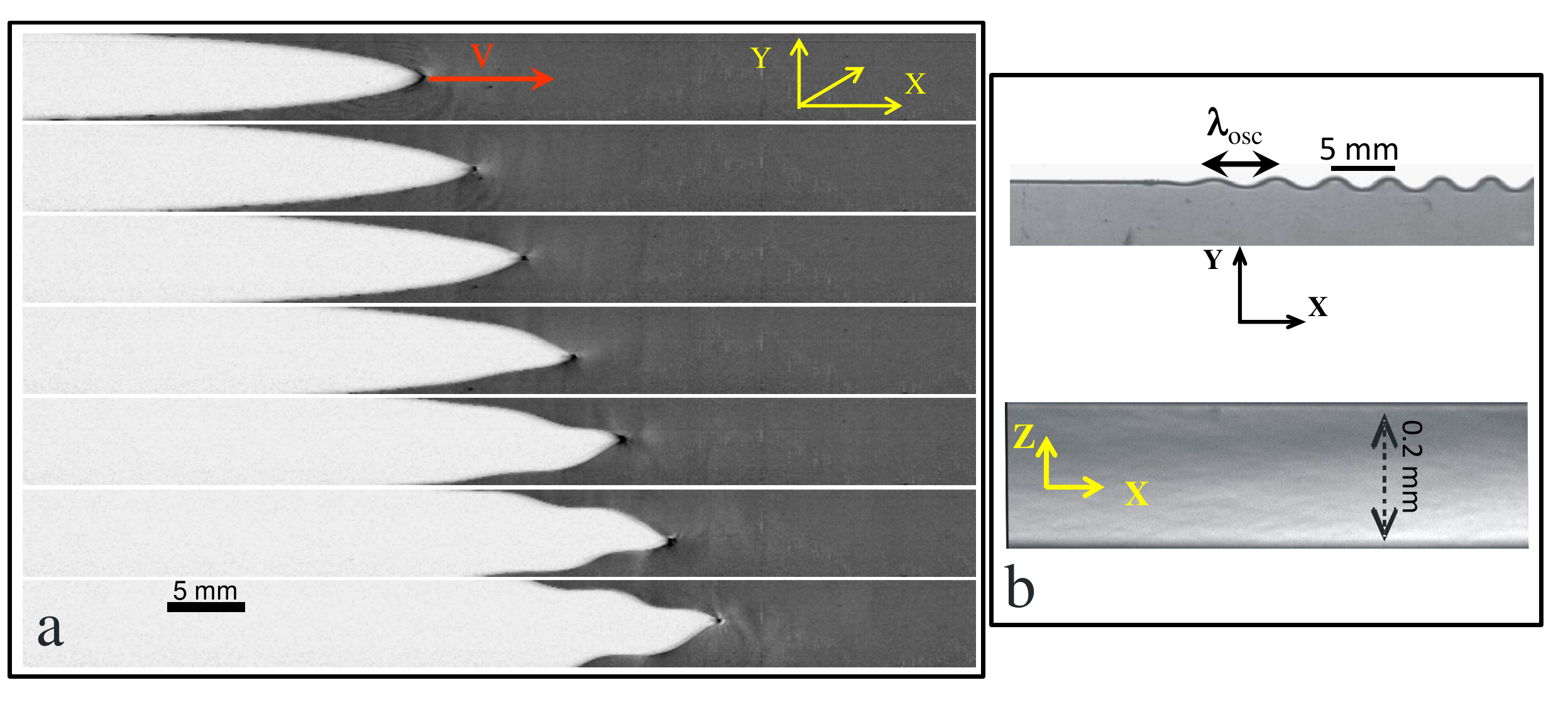}
\caption{When micro-branching is suppressed, a new oscillatory instability occurs at $v\!=\!v_{osc}\!=\!0.9c_R$. (a) Successive $xy$ profiles taken at $690\mu s$ intervals of a crack at the onset of the instability. The tip of an accelerating ``simple" crack starts to oscillate in the $y$ direction, forming a (b) (top) wavy sinusoidal pattern in the $xy$ plane  with a smooth and (bottom) featureless fracture surface.}  \label{Oscillations.pdf}
\end{figure}

While an oscillatory instability has been predicted for rapid cracks \cite{Bouchbinder.07,Henry_Levine.2004}, the predicted wavelength in these models was expected to scale with the size of the system. Experimentally, {\em none} of the characteristic scales (e.g wavelength or amplitude) of the observed oscillations were seen to be at all dependent on sample geometries or dimensions \cite{Livne.07}. It was suggested that this characteristic time/length indicates that a new intrinsic length scale was needed to describe these dynamics. This time/length scale, which could not be explained in the framework of LEFM, hinted that these new effects could be related to the process zone.

\section{Beyond LEFM: regularization of the singularity and dissipation}
\label{beyondLEFM}

As we explained in detail above, LEFM is not a complete and self-contained theory of fracture dynamics. In particular it does not account for near crack tip/front dissipation, consequently it requires the fracture energy as an external input, and it does not account for crack path and velocity selection. There have been numerous attempts to address this problem by formulating models that go beyond LEFM in various aspects. The majority of this work has focused on purely 2D media, though not exclusively. A comprehensive review of these efforts goes well beyond the scope of this paper. For completeness, we briefly mention some of these works below.

A central issue in the context of models of dynamic brittle fracture that go beyond LEFM is how to regularize the LEFM $1/\sqrt{r}$ singularity. Usually this regularization involves dissipation and hence it also gives rise to a fracture energy. A classical way to regularize the continuum LEFM singularity is to introduce a finite-size cohesive zone at the crack tip such that the singularity of the stress field is canceled out on the scale of the cohesive zone \cite{Barenblatt.59;b, Dugdale.60}. Finite Elements Method (FEM) calculations have extended the cohesive zone idea to include cohesive forces between any two elements in the bulk. Cohesive elements calculations are very popular in engineering contexts, and have successfully reproduced several aspects of the dynamics of brittle fracture \cite{Xu.94, Miller.99, Brickstad.80, Ortiz.1999, Ortiz.2001, Belytschko.2002, Molinari.2004, Molinari.2005}. Despite these successes, many aspects of these models are largely phenomenological. These include the introduction of external length scales near the crack tip. In addition, some inherent problems with such models are known to exist -- see \cite{Langer.98, Ching.96b,  Falk.01}.

Much insight has been obtained by the use of lattice models with simple interaction laws, where the lattice constant provides a regularization length scale for the LEFM singularity \cite{Holland.99, Slepyan.02, Marder.93.prl, Marder.95.jmps,Heizler.02,Marder.93.pd,  Ashurst.76, Slepyan.81, Kulakhmetova.84,  Astrom.96, Kaski.96, Cramer.97, Kessler.99, Kessler2000, Kessler.01,  Bernstein.03, Guozden2005, Guozden2006, Ippolito2006, Marder.10}. It is difficult, however, to directly relate these models to the failure of real {\em non-crystalline} solids, where a natural regularization scale is not obviously apparent and well-defined slip systems do not exist. Recently, a continuous random network model has been developed and studied in detail \cite{Heizler.11}. Direct molecular dynamics simulations have also successfully reproduced some of the phenomenology of dynamic brittle fracture \cite{Holland.99,Abraham.94,Gumbsch.97,Gumbsch.01,Rountree.02,Kalia.03}. Like lattice models, these calculations are generally performed in an ordered atomic material using largely phenomenological potentials. Indeed, attempts to use realistic crystalline potentials (e.g. in the fracture of silicon) have failed to quantitatively reproduce experiments \cite{Hauch.99,Holland.er.98}. Later on, it has been shown that quantum mechanical contributions to these potentials can not be ignored \cite{bernstein.09,kermode.08}.

Diffuse interface phase-field models provide a methodology to regularize the near-tip singularity. These models are self-consistent continuum formulations of brittle crack problems that incorporate an auxiliary phenomenological ``phase field" to link the near-tip behavior to the linear elastic fields far from the tip \cite{aranson2000continuum,karma2001phase,eastgate2002fracture,marconi2005diffuse}. These models provide a useful mathematical machinery that allows a self-consistent selection of the crack's speed, tip/front shape and path \cite{Kessler.01,Karma.01, Aranson.00,Karma2004, Spatschek.06, spatschek2007phase, spatschek2011phase}. In particular, they introduce a regularization length scale for the LEFM singularity and a time scale for near-tip dissipation. Phase transformation models of fracture were also studied in detail \cite{Pilipenko2007, Fleck2011}. While largely phenomenological at present, this class of models is promising and has been useful in elucidating several important aspects of crack dynamics. These include laws of crack motion in isotropic and anisotropic quasi-static 2D fracture \cite{hakim2005crack, Hakim.09} and the evolution of complex quasi-static crack patterns in 3D under mixed-mode loading \cite{Karma.10,leblond2011theoretical}.

Many of the models that go beyond LEFM have addressed the issue of crack branching. As we stressed above, the vast majority of these focused on 2D where a crack tip bifurcates into two tips (``macro-branching''). These include lattice models \cite{Marder.95.jmps, Sander.99, Marder.93.pd}, a continuous random network model \cite{Heizler.11}, molecular dynamics simulations \cite{Holland.99, Abraham.94}, phase-field models \cite{Kessler.01,Karma.01, Aranson.00,Karma2004, Spatschek.06, spatschek2007phase, spatschek2011phase}, phase transformation models \cite{Pilipenko2007, Fleck2011} and cohesive zone formulations coupled with finite element calculations \cite{Miller.99, Molinari.2004, Molinari.2005}. The intrinsic 3D nature of the instability, however, has until now escaped our theoretical understanding \cite{Karma.10, leblond2011theoretical, Adda-Bedia.13}. While a number of explanations for the 2D instability have been proposed \cite{Gao.93,Gao.96, Buehler.03,Bouchbinder.05a,Adda-Bedia.05,Pilipenko2007}, the precise physical mechanism leading to this instability has not yet been established.

Finally, the role of elastic nonlinearities in dynamic fracture has also been the subject of recent research \cite{Gao.93,Gao.96,Buehler.03,Buehler2006,Livne.2010,Livne.08,Bouchbinder.08a,Bouchbinder.09, Bouchbinder.08b, Harpaz.2012}. The basic idea is that the LEFM singularity implies the concentration of large deformations near the crack tip that must invalidate the assumption of a linear elastic behavior. On quite general grounds one expects linear elasticity to break down in favor of nonlinear elasticity before irreversible processes set in. Based on this idea, a weakly nonlinear elastic theory of dynamic fracture has been developed. This will be discussed in detail in the next section.

\section{A weakly nonlinear theory of fracture: The effects of nonlinear elasticity near the crack tip} \label{WNLsection}

In section \ref{instabilities}, we described two different instabilities that arise in brittle fracture. Neither of these can be understood in the framework of the fracture dynamics that are embodied in LEFM. We have also shown that, as long as a ``simple'' crack remains stable, crack dynamics are in excellent quantitative agreement with LEFM, as demonstrated in figure \ref{lefm_comparison}.

What is going on? Energy balance is at the heart of the single crack dynamics. LEFM uses an assumption of linear elasticity (nearly) everywhere to calculate the flux of energy transported from the external loading by means of the linear elastic fields (where materials do not undergo large deformations) to the region of a crack's tip. On the one hand, the highly deformed regions that exist in the near vicinity of a crack's tip can not be dealt with in the framework of LEFM. On the other hand, as long as these highly deformed regions are ``passive" and exhibit no important dynamics of their own, they can be incorporated into the assumption of small-scale yielding and crack dynamics can be completely described in the LEFM framework. The unexplained ``intrinsic" scale that lies at the heart of the oscillatory instability, described in the previous section, however, provided a hint that something might be fundamentally missing in this assumption. An equation of motion for a crack should determine both the crack growth rate and its direction of propagation \cite{Adda-Bedia.99,  Hakim.09, Oleaga.01, Bouchbinder.09b}. It is crucial to understand that energy balance, which is behind the crack growth rate equations (\ref{eqn_motion}) and (\ref{marder}), provides us with no information whatsoever about the direction of crack propagation, only about the speed of propagation. Without an equation that determines a crack's path, the question of path stability cannot even be formulated mathematically. Intrinsic length scales might be at the heart of such an equation.

The $K/\sqrt{r}$ scaling of the stress field predicted by LEFM as the crack tip is approached is a power-law, which possesses no intrinsic scale. That is, our fundamental understanding of fracture is based on the assumption that fracture dynamics are governed by the singular stresses that dominate all contributions to the stress field up to the (``single") point where the material is broken. All of the details of how a given system is loaded and what the geometry of the sample and the crack is, are incorporated into a single scalar quantity, the stress intensity factor, $K$. This is sometimes called the assumption of ``autonomy", which means that the mechanical state within the near-tip region is uniquely determined by the value of $K$ and is otherwise independent of the applied loadings
and the geometric configuration in a given problem \cite{Broberg.99}. In particular, the length scale inherited in $K$ (whose dimension is stress times square root of a length) is extrinsic, associated with either the geometry or the loading of the problem. Were this framework capable of explaining the oscillatory instability discussed above, it would have inevitably implied that the wavelength of oscillations is determined by an extrinsic length scale. As the experiments clearly show that this is not the case, we must therefore look for new physics and a new length scale in the region where the assumption of {\em linear} elasticity breaks down, i.e. in the near-tip region.

The use of soft gels provides us with a unique experimental opportunity to explore this elusive region. The slow propagation velocities of dynamic cracks within these materials have enabled us to perform measurements of unprecedented accuracy of the material deformations in the near-tip vicinity of truly dynamic cracks. This allows us to quantitatively examine the form of the deformation fields at scales surrounding the crack tip where the assumption of linear elastic behavior breaks down. Before we discuss in detail the outcome of these experimental investigations, we will first describe a new theoretical framework that goes beyond LEFM. This will set the stage for the quantitative analysis to follow.

As our goal is to go beyond LEFM, we should ask ourselves what is the first physical process that intervenes when LEFM breaks down near a crack's tip. LEFM is based on the assumption of {\em linear} reversible (elastic) deformation. While traditional approaches focus on irreversible deformation near crack tips, e.g. plastic deformation \cite{hutchinson.68, rice.rosengren.68}, our premise is that {\em first} linearity breaks down, while the deformation remains reversible. In a nutshell, we maintain that the harmonic (quadratic) approximation for the deviation from a stress-free configuration -- which is at the heart of LEFM as in equation (\ref{U_varepsilon}) -- must first give way to anharmonic (nonlinear) corrections that do not necessarily involve irreversible processes (e.g. particles rearrangements, decohesion etc.). Below we explore the theoretical implications of this idea.

\subsection{Finite deformations: a tutorial}
\label{preliminary}

To set the stage for theoretical developments based on nonlinear elasticity, we will need some background in the theory of finite elastic deformations. This theory is well-developed and is summarized in various textbooks, see for example \cite{Holzapfel, 51Murnaghan}. To render this paper as self-contained as possible, we briefly review some basic elements of this theory. We start with equations (\ref{phi})-(\ref{disp_grad}) and define the deformation gradient $\B F$ as $\B F \!=\! \nabla \B \phi \!=\! \B I \!+\! \B H$.
%\begin{equation}
%\label{F}
%\B F = \nabla \B \phi = \B I + \B H \ ,
%\end{equation}
%where $\B I$ is the identity tensor.
%Note that since $\B H$ is not symmetric, so is $\B F$.
The Green-Lagrange tensor $\B E$ is defined as
\begin{equation}
\label{E_strain}
\B E = \frac{1}{2}(\B H + \B H^T + \B H^T \B H) =\B \varepsilon + \frac{1}{2}\B H^T \B H \ ,
\end{equation}
where the linear strain tensor $\B \varepsilon$ is given in equation (\ref{infinitesimal_strain}). $\B E$ is a rotationally invariant tensor that measures the change in length of material elements and is evidently nonlinear in $\B H$. The nonlinear elastic energy density functional $U$, i.e. the elastic energy per unit reference volume, of isotropic materials can be expressed as
\begin{equation}
U = U(I_{\B E}, II_{\B E}, III_{\B E}) \ ,
\end{equation}
where $\{ I_{\B E}, II_{\B E}, III_{\B E} \}$ is a set of invariants which we take here to be the principal moments
\begin{equation}
\label{invar}
{I}_{\B E} \equiv tr\!\B E,\quad {II}_{\B E} \equiv tr\!\B E^2,\quad {III}_{\B E} \equiv tr\!\B E^3 \ .
\end{equation}

The first Piola-Kirchhoff stress tensor $\B s$, that is work-conjugate to the deformation gradient $\B F$, is defined as
\begin{equation}
\label{1st_PK}
\B s = \frac{\pa U}{\pa \B F} = \B F \frac{\pa U}{\pa \B E} \ .
\end{equation}
$\B s$ quantifies forces in the deformed configuration per unit area in the reference configuration. As will be shown below, this quantity is very useful in calculations.
The Cauchy stress tensor $\B \sigma$, which is the true mechanical stress that appears in momentum balance equations, can be expressed in terms of $\B s$ as
\begin{equation}
\label{Cauchy}
\B \sigma = \frac{\B s \B F^T}{det \B F} \ .
\end{equation}
The linear momentum balance equation is formulated in the deformed configuration in terms of the Cauchy stress $\B \sigma$, cf. equation (\ref{EOM}). The differential operators are understood to be defined with respect to the deformed coordinates $\B x'$, even though in the context of linear elasticity in which these equations were introduced, we made no distinction between the reference and deformed coordinates. The momentum balance equation can be rewritten in terms of the first Piola-Kirchhoff stress tensor $\B s$ and the reference (undeformed) coordinates $\B x$ as
\begin{equation}
\label{EOM_s}
\nabla_{\B x} \cdot \B s(\B x,t) = \rho_0(\B x) \pa_{tt}\B u(\B x,t)  \ ,
\end{equation}
where $\rho_0(\B x)$ is the (time-independent) mass density in the reference configuration. Angular momentum balance, $\B \sigma\!=\!\B\sigma^T$, can be expressed as
\begin{equation}
\label{angular}
\B s \B F^T = \B F \B s^T  \ .
\end{equation}
It is important to note that equation (\ref{EOM_s}) is defined with respect to a {\em fixed, known domain}. This is a great simplification when compared to equation (\ref{EOM}), which is defined with respect to an {\em evolving, yet unknown domain}. This feature makes the first Piola-Kirchhoff stress tensor $\B s$ of equation (\ref{1st_PK}) very useful, especially for the application of these equations to fracture mechanics, where a crack introduces time-dependent traction-free boundary conditions described mathematically as moving branch-cuts in the undeformed configuration. Equation (\ref{EOM_s}) will serve as a starting point for deriving the weakly nonlinear theory.

\subsection{The weakly nonlinear theory of dynamic fracture}
\label{WNFM}

The weakly nonlinear theory of dynamic fracture incorporates (weak) elastic nonlinearity by expanding the equations of motion describing the deformation in a medium containing a propagating crack up to second order in the displacement-gradient tensor $\B H$, which is regarded as the small parameter in the problem. The starting point in the derivation is to write down the elastic energy density $U$ up to $\C O(\B E^3)$ as
\begin{equation}
\label{U_E}
\hspace{-2cm} U(\B E) = \frac{1}{2} \lambda \left(tr\!\B E\right)^2 + \mu ~tr\!\B E^2 + \beta_1 \left(tr\!\B E\right)^3 + \beta_2 tr\!\B E~ tr\!\B E^2 + \beta_3 tr\!\B E^3 + \C O(\B E^4) \ ,
\end{equation}
where $\lambda$ and $\mu$ are the standard Lam\'e constants and $\{\beta_1, \beta_2, \beta_3\}$ are the second order elastic constants. The second order elastic constants are basic physical quantities that represent the leading anharmonic contributions to the interatomic interaction potential. These anharmonic contributions are known to be the origin of many important physical properties and effects such as the Gr\"{u}neisen parameters, deviations from the Dulong-Petit law at high temperatures, thermal expansion and the existence of thermal resistance; their implications to dynamic fracture are discussed in this review. $\{\beta_1, \beta_2, \beta_3\}$ are, therefore, not free parameters, but rather quantities that are either calculated from a fully nonlinear elastic energy functional, if known, or measured directly in experiments (see section \ref{2nd_constants} below). Equation (\ref{U_E}) reduces to the linear elastic energy density equation (\ref{U_varepsilon}) under two conditions; the strain measure in equation (\ref{E_strain}) should be linearized with respect to $\B H$ and the second order elastic constants should be set to zero.

Like the first order elastic constants, where the pair $\{\lambda, \mu\}$ can be replaced by equivalent pairs (e.g. the Young's and bulk moduli), the choice of second order elastic constants is not unique. For example, the Murnaghan coefficients $\{\ell, m, n\}$ constitute an alternative set of second order constants \cite{51Murnaghan}. They are simply related to $\{\beta_1, \beta_2, \beta_3\}$ defined above through
\begin{equation}
\label{Murnaghan}
\ell\!=\!3\beta_1+\beta_2,~ m\!=\!\beta_2+3\beta_3/2, ~ n\!=\!3\beta_3 \ .
\end{equation}

It is important to note that $U(\B E)$ in equation (\ref{U_E}) is valid for a general isotropic material in 3D, where $\B E$ is regarded as 3D tensor. In many cases, however, one is interested in situations in which the deformation state is 2D due to some simplifying physical conditions/assumptions. The only change to the above formalism would be in redefining the second order elastic constants $\{\beta_1, \beta_2, \beta_3\}$ in equation (\ref{U_E}). For example, under 2D plane-strain conditions \cite{Landau.86} we have
\begin{equation}
\label{E_ps}
\B E = \begin{pmatrix} E_{xx} \; E_{xy} \; 0 \\ E_{yx} \; E_{yy} \; 0 \\ ~~0 \;\;\; ~0\;\;\; 0 \end{pmatrix} \ ,
\end{equation}
for which equation (\ref{U_E}) becomes
\begin{equation}
\label{U_E_plane_strain}
 U^{2D} = \frac{1}{2} \lambda \left(tr\!\B E_{2D}\right)^2 + \mu ~tr\!\B E_{2D}^2 + \left(\beta_1+\frac{\beta_2}{3}\right) \left(tr\!\B E_{2D}\right)^3 + \left(\beta_3+\frac{2\beta_2}{3}\right) tr\!\B E_{2D}^3 \ ,
\end{equation}
where $\B E_{2D}$ is a 2D tensor composed of the non-vanishing elements of $\B E$ in equation (\ref{E_ps}). To keep things as general as possible, we use the general 3D form of equation (\ref{U_E}) to derive general results (i.e. not for a specific material or geometry) below, even when dealing with 2D deformation conditions. This simply means that when applying these results to a specific 2D problem, the second order elastic constants $\{\beta_1, \beta_2, \beta_3\}$ should be properly selected. For example, in the case of the plane-strain conditions of equation (\ref{U_E_plane_strain}), we define $\{\tilde\beta_1\!=\!\beta_1+\beta_2/3, \tilde\beta_2\!=\!0, \tilde\beta_3\!=\!\beta_3+2\beta_2/3\}$ and use these instead of $\{\beta_1, \beta_2, \beta_3\}$ in all of the results to follow. Another example, for plane-stress conditions, will be discussed in section \ref{comparison}.

To proceed, we use equation (\ref{U_E}) in equation (\ref{1st_PK}) and expand the result up to second order in $\B H$ to obtain
\begin{eqnarray}
\label{s_2nd_order}
\B s = (\B I + \B H)\frac{\pa U}{\pa \B E} \simeq \lambda~\! tr\B \varepsilon \B I+ 2 \mu \B \varepsilon + \\
\frac{1}{2} \lambda~ tr(\B H^T\B H) \B I + \mu\, \B H^T \B H+ \lambda \,tr \B \varepsilon \B H + 2 \mu\, \B H \B \varepsilon +\nonumber\\
3 \beta_1\, (tr\B \varepsilon)^2 \B I + \beta_2~\! tr\B \varepsilon^2 \B I + 2 \beta_2 ~\!tr\B \varepsilon~ \B \varepsilon + 3\beta_3\, \B \varepsilon^2 + \C O(\B H^3) \equiv\B s^{1st} + \B s^{2nd} \ .\nonumber
\end{eqnarray}
Here $\B s^{1st}$ stands for the part of $\B s$ that is linear in $\B H$ and $\B s^{2nd}$ stands for the part of $\B s$ that is quadratic in $\B H$. Related second order stress tensors appeared in the literature, see for example \cite{51Murnaghan, 30Sig, 65TN}. It is straightforward to show that the last expression automatically satisfies the angular momentum balance in equation (\ref{angular}) to $\C O(\B H^2)$. Then, $\B s$ of equation (\ref{s_2nd_order}) can be used in the linear momentum balance in equation (\ref{EOM_s}) to obtain the most general equations of motion for the displacement field $\B u$, consistent to second order in $\B H$.

Equation (\ref{s_2nd_order}) reveals the structure and origin of second order nonlinearities. The first two terms are linear in $\B H$ and correspond to the usual Hooke's law for small deformations. The remaining terms are second order in $\B H$. These terms can be classified into two different types. The first type corresponds to second order nonlinearities that emerge from the fact that the basic rotationally invariant strain measure $\B E$ is quadratic in $\B H$, cf. equation (\ref{E_strain}). These second order nonlinearities do not involve new constitutive parameters (hence sometimes termed ``geometric nonlinearities'') and can be readily identified as the nonlinear terms in equation (\ref{s_2nd_order}) that are proportional to the Lam\'e constants $\lambda$ and $\mu$. The remaining nonlinear terms in equation (\ref{s_2nd_order}) emerge from the fact that the constitutive relation is nonlinear (hence sometimes termed ``constitutive nonlinearities'') and therefore are proportional to the second order elastic constants $\{\beta_1, \beta_2, \beta_3\}$. In the most general situation, we expect both types of nonlinearities to be relevant.

We are now ready to derive the most general weakly nonlinear equations for the displacement field $\B u$ and to apply these to an asymptotic expansion near the edge of a crack, where displacement-gradients become too large for the linear approximation to be accurate. To this end, we introduce a controlled expansion of $\B u$ of the form
\begin{equation}
\label{expansion}
\B u \simeq \epsilon\, \tilde{\B u}^{(1)}+ \epsilon^2 \tilde{\B u}^{(2)} + \C O(\epsilon^3) \equiv \B u^{(1)}+  \B u^{(2)}\ ,
\end{equation}
where $\epsilon$ is a measure of the magnitude of displacement-gradients, not to be confused with the small-deformation strain tensor $\B \varepsilon$ defined in equation (\ref{infinitesimal_strain}). The expansion in equation (\ref{expansion}) can be now substituted into the expression for $\B s$ in equation (\ref{s_2nd_order}) and then the equations of motion (\ref{EOM_s}) can be expanded in orders of $\epsilon$.
To first order in $\epsilon$ we obtain the standard Lam\'e equation, cf. equation (\ref{Lame}).

The second order equation in $\epsilon$ takes the form
\begin{equation}
\mu\nabla^2{\B u^{(2)}}+(\lambda+\mu)\nabla(\nabla\cdot{\B u^{(2)}})+
\B {\C F}[\B u^{(1)}]=\rho_0\pa_{tt}{\B u}^{(2)}\ ,
\label{secondO}
\end{equation}
where $\B{\C F}$ is a functional whose components are given by
\begin{eqnarray}
\C F_x[\B u^{(1)}] &=& \pa_x s^{2nd}_{xx}[\B u^{(1)}] + \pa_y s^{2nd}_{xy}[\B u^{(1)}],\nonumber\\
\C F_y[\B u^{(1)}] &=& \pa_x s^{2nd}_{yx}[\B u^{(1)}] + \pa_y s^{2nd}_{yy}[\B u^{(1)}] \ .
\label{Cal_F}
\end{eqnarray}
Equation (\ref{secondO}) has the form of the Lam\'e equation (equation (\ref{Lame})) with
an added effective body loading given by $\B{\C F}\!\!\left[\B u^{(1)}\right]$. The expression for $\B s^{2nd}$, defined in equation (\ref{s_2nd_order}), is too lengthy to be presented here. Note that we already focus on 2D here\footnote{In the spirit of equation (\ref{expansion}), the stress $\B s$ can be also expanded in powers of $\epsilon$ according to $\B s\!\simeq\!\epsilon\, \tilde{\B s}^{(1)}+ \epsilon^2 \tilde{\B s}^{(2)} + \C O(\epsilon^3) \!\equiv\! \B s^{(1)}+  \B s^{(2)}$. Therefore, equation (\ref{secondO}) is in fact of the form $\pa_j s^{(2)}_{ij} \!=\! \rho_0 \pa_{tt}{u}^{(2)}_i$. It is important to note the difference between $\B s^{2nd}$ defined in equation (\ref{s_2nd_order}) and $\B s^{(2)}$. The former is the second order contribution to $\B s$ in terms of $\B H$, while the latter is the second order contribution to $\B s$ in terms of $\epsilon$.}.

To apply equation (\ref{Cal_F}) to the dynamics of cracks, we recall that a crack is defined as two surfaces that cannot support stresses and therefore is characterized by traction-free boundary conditions on its faces $\sigma_{ij}\hat n'_j\!=\! 0$, where $\hat{\B n}'$ is the outward normal on the actual crack faces, i.e. in the deformed configuration.
These boundary conditions can be rewritten in the undeformed configuration in terms of the first Piola-Kirchhoff stress tensor $\B s$ as $s_{ij}\hat{n}_j \!=\!0$, where $\hat{\B n}$ is the normal to the crack faces in the reference configuration. The latter introduces a great technical simplification as they are defined with respect to a known domain defined by the crack faces in the undeformed configuration. Note that since we will not be interested below in solving a global boundary value problem, but rather focus on a near crack edge asymptotic analysis, we do not explicitly consider the external boundary conditions imposed on the sample; the latter are needed to render the global boundary value problem well posed.

The boundary conditions $s_{ij}\hat{n}_j \!=\!0$ can be expressed explicitly for a propagating crack as
\begin{eqnarray}
\label{BC_s_explicit}
s_{xy}(r,\varphi\!=\!\pm\pi)=s_{yy}(r,\varphi\!=\!\pm\pi)= 0 \ .
\end{eqnarray}
The boundary conditions of equation (\ref{BC_s_explicit}), considered to first order in $\epsilon$, yield equation (\ref{BC1st}). To second order in $\epsilon$, these boundary conditions read
\begin{eqnarray}
&& \label{BC2nd} -\mu\,r^{-1}\pa_\varphi u_x^{(2)}-\mu\pa_ru_y^{(2)}-\C S_x[\B u^{(1)}]=0\ , \nonumber\\
&& -(\lambda+ 2\mu)r^{-1}\pa_\varphi u_y^{(2)}-\lambda \,\pa_r u_x^{(2)}-\C S_y[\B u^{(1)}]=0 \ ,
\end{eqnarray}
for $\varphi=\pm\pi$ \footnote{Note that in the spirit of the previous footnote, these boundary conditions are equivalent to $s^{(2)}_{xy}(r,\varphi\!=\!\pm\pi)\!=\!0$ and $s^{(2)}_{yy}(r,\varphi\!=\!\pm\pi)\!=\!0$, respectively.}. $\B{\C S}[\B u^{(1)}]$ is an effective surface force vector, which is quadratic in $\B u^{(1)}$, and whose components are given by
\begin{eqnarray}
\C S_x[\B u^{(1)}] &=& -s^{2nd}_{xy}(r,\varphi\!=\!\pm\pi) \ ,\nonumber\\
\C S_y[\B u^{(1)}] &=& -s^{2nd}_{yy}(r,\varphi\!=\!\pm\pi) \ ,
\label{Cal_S}
\end{eqnarray}
where $\B s^{2nd}$ is defined in equation (\ref{s_2nd_order}).

Our goal here is to consider the order $\epsilon^2$ problem for the mode I (tensile) symmetry of fracture. The theory for mode II (shear) symmetry is presented in detail in \cite{Harpaz.2012}. The second order problem is defined by equations (\ref{secondO}) and (\ref{BC2nd}). Using the leading terms (proportional to $K_I$) in the first order displacement field $\B u^{(1)}$ of equations (\ref{u_1st_I}) in equations (\ref{Cal_F}) and (\ref{Cal_S}), we obtain
\begin{eqnarray}
\label{F_S}
\B{\C F}(r,\varphi) &=& \frac{K^2_I \B g(\varphi;v)}{32\pi\mu^2 r^2}\ , \nonumber\\
\C S_x(r,\pm\pi) &=& 0 \ ,\nonumber\\
\C S_y(r,\pm\pi) &=&  \frac{K^2_I \kappa(v)}{32\pi \mu^2 r} \ ,
\end{eqnarray}
where $\B g(\varphi;v)$ and $\kappa(v)$ have the dimension of stress and the numerical factor $32\pi$ is introduced in order to be consistent with previously used definitions \cite{Bouchbinder.08a, Bouchbinder.09, 09Bouchbinder_a}. $\kappa(v)$ is given explicitly as
\begin{equation}
\label{kappa}
\kappa(v) = -\frac{16\alpha_d^2 v^4 \left(\lambda + \mu \right)}{c_s^4 D^2(v)} \ ,
\end{equation}
which happens to be independent of the second order elastic constants $\{\beta_1, \beta_2, \beta_3\}$. The vector function $\B g(\varphi;v)$ is too long to be reported, but it possesses the following symmetry properties
$g_x(\varphi;v)\!=\!g_x(-\varphi;v)$ and $g_y(\varphi;v)\!=\!-g_y(-\varphi;v)$. We emphasize that these functions depend on material properties through the elastic constants, though we do not explicitly write this dependence in the functions' arguments.

Together with the vanishing effective surface force component $\C S_x$ and $\C S_y(r,\pi)\!=\!\C S_y(r,-\pi)$ in equation (\ref{F_S}), we see that the second order problem corresponds to a mode I linear elastic crack problem with an effective body force $\B{\C F}(r,\varphi)$ that scales as $r^{-2}$ and an effective surface force $\C S_y(r,\pm\pi)$ that scales as $r^{-1}$, both are quadratic in $K_I$.

The solution of equation (\ref{secondO}) was derived in detail in \cite{Bouchbinder.08a, Bouchbinder.09} and takes the form
\begin{eqnarray}
\label{solution}
\hspace{-2.5cm} u_x^{(2)}\!\!&=&\!\!\frac{K^2_I}{32\pi\mu^2}\!\!\left[\!A\log{r}\!+\!\frac{A}{2}\log{\left(\!\!1\!-\!\frac{v^2\sin^2\!\!\varphi}{c_d^2}\!\right)}\!+\!B\alpha_s\log{r}\!+\!\frac{B \alpha_s}{2}\log{\!\!\left(\!\!1\!-\!\frac{v^2\sin^2\!\!\varphi}{c_s^2}\!\right)}\!+\!\Upsilon_x(\varphi;v)\!\right],\nonumber\\
\hspace{-2.5cm} u_y^{(2)}&=&\frac{K^2_I}{32\pi\mu^2}\left[-A\alpha_d\varphi_d-B\varphi_s+\Upsilon_y(\varphi;v)\right] \ .
\end{eqnarray}
The displacement-gradients derived from this solution
\begin{equation}
\label{1/r}
\nabla {\B u}^{(2)} \!\propto\! 1/r \ ,
\end{equation}
all exhibit a singularity that is stronger than the $1/\sqrt{r}$ singularity of LEFM.  In addition, this solution is characterized by $\log{r}$ displacement contribution. Both features were directly confirmed in experiments \cite{Bouchbinder.08a, Bouchbinder.09, 09Bouchbinder_a}. $\B \Upsilon(\varphi;v)$ is an $r$-independent solution of equation (\ref{secondO}) which does not satisfy the boundary conditions of equations (\ref{BC2nd}). The Fourier series representation of $\B \Upsilon(\varphi;v)$ reads
\begin{equation}
\Upsilon_x(\varphi;v) \simeq \sum_{n=1}^N c_n(v) \cos(n\varphi)\quad\hbox{and}\quad \Upsilon_y(\varphi;v) \simeq \sum_{n=1}^N d_n(v) \sin(n\varphi) \ .
\end{equation}
Specific solutions demonstrated that a small $N$ is sufficient to obtain accurate representations even at very high crack propagation speeds \cite{Bouchbinder.08a}.

As $\B \Upsilon(\varphi;v)$ does not satisfy the boundary conditions of equations (\ref{BC2nd}), it does not constitute a complete solution of the second order problem. The remaining part, cf. equation (\ref{solution}), comes from a solution of the homogeneous counterpart of equation (\ref{secondO}) (i.e. the standard Lam\'e equation) that is characterized by a $1/r$ displacement-gradients singularity. This property is needed in order to be able to satisfy the second boundary condition in (\ref{BC2nd}), which contains an effective surface force that scales as $1/r$. Substituting equation (\ref{solution}) into this boundary condition results in the following relation between $A$ and $B$
\begin{equation}
A = \frac{2\mu B \alpha_s -(\lambda+2\mu)\pa_\varphi \Upsilon_y(\pi;v)-\kappa(v)}{\lambda - (\lambda+2\mu)\alpha_d^2} \ .
\label{A_B_bc}
\end{equation}
This relation reduces to the one reported in \cite{Bouchbinder.08a}, where $\lambda\!=\!2\mu$ and $c_d\!=\!2c_s$ were used for a neo-Hookean material, and $\kappa(v)$ was measured in units of $\mu$.

The parameter $B$ seems to remain undetermined in the
solution in equation (\ref{solution}). If true, this result is remarkable
as it suggests that the concept of the autonomy of the near crack
tip nonlinear region (see the discussion at the beginning of section \ref{WNLsection} and \cite{98Fre, Broberg.99}) is not always valid. However, autonomy is a central concept in fracture
mechanics \cite{98Fre,Broberg.99} and it is difficult to see how
it could be violated. How, then, can one reconcile the fact that equation
(\ref{solution}), with equation (\ref{A_B_bc}), satisfies the second
order asymptotic boundary-value problem for {\em all} $B$,
but still be consistent with the concept of autonomy? The resolution to this apparent paradox was discussed in detail in \cite{Bouchbinder.09, 09Bouchbinder_a}. It was shown that the $1/r$ singularity is unique in the sense that it gives rise to a spurious resultant (integrated) force in the crack's parallel direction (where no boundary conditions are imposed) and hence the resultant force balance is not automatically satisfied in this direction, even though the solution itself satisfies the asymptotic boundary-value problem. Since such an unbalanced/spurious force is unphysical (as no physical process, except for inertial motion, can produce such a net force), it must be eliminated by demanding that the resultant (integrated) Newton's equation (both per unit sample thickness)
\begin{equation}
\label{f_dynamic}
f_i \equiv \int_{-\pi}^{\pi} s_{ij} n_j r d\varphi = v^2 \rho_0 \int_0^r \tilde{r} d\tilde{r} \int_{-\pi}^{\pi} \pa_{xx} u_i d\varphi \equiv \dot{p}_i
\end{equation}
is satisfied. Here $\B f$ is the net force per unit sample thickness acting on a line of radius $r$ encircling a crack's tip and $\dot{\B p}$ is the time rate of change of momentum per unit sample thickness of the material enclosed in the circle. Note that the steady state relation $\pa_t\!=\!-v\pa_x$ was used.
The y-component of equation (\ref{f_dynamic}) is automatically satisfied due to the mode I symmetry. The x-component imposes a real physical constraint. It was further shown in \cite{09Bouchbinder_a} that unlike the $1/\sqrt{r}$ fields of LEFM (derived from equations (\ref{u_1st_I})), the $1/r$ singular fields carry no net momentum rate,
\begin{equation}
\label{dot_p}
\dot p_x^{(2)} \equiv v^2 \rho_0 \int_0^r \tilde{r} d\tilde{r} \int_{-\pi}^{\pi} \pa_{xx} u_x^{(2)} d\varphi = 0 \ .
\end{equation}
This result implies that
\begin{equation}
\label{dynamic_condition}
f^{(2)}_x\equiv\int_{-\pi}^{\pi} s^{(2)}_{xj} n_j r d\varphi = 0 \ .
\end{equation}
The latter constraint allows the determination of $B$. Therefore, $B$ is uniquely determined once the LEFM asymptotic fields are known and the autonomy of the near tip region is retained. An example of calculating $B$ was given in \cite{09Bouchbinder_a}.

Before we turn to discussing experimental tests of this theory, we discuss some important issues related to it.

\subsection{The determination of the second order elastic constants}
\label{2nd_constants}

An essential input to the weakly nonlinear theory is the set of second order elastic constants $\{\beta_1, \beta_2, \beta_3\}$ that were defined in equation (\ref{U_E}).  As discussed above, the second order elastic constants are directly related to the leading anharmonic contributions to the interatomic interaction potential that are known to be the origin of many important physical effects. We include this section since we believe that the topic of second order elastic constants is not well-known to the general reader. To both rectify this and to emphasize the fact that these are real physical constants that lead to real physical effects, we will now briefly review a number of methods that have been successfully used to measure these quantities.

Methods for measuring the second order elastic constants are well-developed and rather widely used. Nevertheless, and in contrast to the first order elastic constants $\lambda$ and $\mu$, their measured values for many materials are not well-documented and hence usually are not easily accessible. Here we briefly review some of the experimental and theoretical methods used for obtaining the second order constants.

A direct theoretical method for determining the second order elastic constants becomes available once first-principles calculations of the structure and the interatomic interaction potentials of a given material are known. In that case, the elastic energy functional of equation (\ref{U_E}) can be directly calculated and the elastic constants are obtained from a polynomial fit to the calculated energy-strain relation. In fact, yet higher order elastic constants can be obtained using this procedure. Recent first-principles density-functional theory calculations explicitly demonstrated this approach for single crystals \cite{07ZWG, 09WL}. Additional recent work employed tight-binding atomistic simulations to calculate the second order elastic constants of monolayer graphene \cite{08LWKH,09CPGC}. Naturally, such first-principles approaches are more adequate for crystals with a well-defined symmetry than for amorphous solids.

Some of the experimental methods for determining the second order elastic constants are described in the 1981 review paper of Hiki \cite{81Hiki}, where a list of papers in which available data (as of 1981) is compiled. Static methods mainly involve measuring the onset of nonlinear variations of stress vs. deformation. Such static measurements for many materials have already been carried out in \cite{29Bridgman}. In addition to the static methods, dynamic methods exist that are mainly based on wave propagation. One such method \cite{81Hiki,68GB,73YB} is based on the idea that in a stress-free material, finite
amplitude waves of a given fundamental frequency generate higher
order harmonics (waves whose frequencies are integer
multiples of the fundamental frequency) due to elastic
nonlinearities. Second order elastic nonlinearities give rise to the generation of second harmonics whose amplitude depends on the second order elastic constants. This method has been applied to a number of  materials, see for example \cite{81Hiki, 68GB, 73YB}.

The most widely used method for measuring the second order elastic constants is acoustoelasticity, which is the acoustical analog of photoelasticity in optics. The method is based on measuring the speed of small amplitude plane waves that are superimposed on an applied static stress. The first expressions for the wave speeds as a function of the applied stress and the resulting second order elastic constants appeared in the pioneering
work of Hughes and Kelly \cite{53HK}. Additional acoustoelastic techniques are discussed by Crecraft \cite{Crecraft}, who developed a rather accurate acoustoelastic technique based on ultrasonic waves.

Acoustoelasticity has been extensively applied to many materials, employing various techniques for measuring the wave speeds. For example, a Brillouin spectroscopy method was developed and applied to measure the second order elastic coefficients of solid polymers \cite{91KGSZD}, soda-lime-silica glass \cite{99CLVVL} and a bulk metallic glass \cite{07KKK}. Recently, a coda wave interferometry method was used to determine
the second order elastic constants of complex solids such as concrete \cite{09PGM}.

Finally, the second order elastic constants can be determined by expanding a well-established fully nonlinear elastic energy functional. Such an energy functional usually becomes available by directly fitting
experimental data for highly compliant materials, e.g. rubber-like materials and elastomer gels. In some cases the form of the energy functional is supported by a microscopic model, e.g. the neo-Hookean
model, but in other cases symmetry considerations and pure phenomenology are sufficient, e.g. the Blatz-Ko energy functional that was used to fit some experimental data for foam rubber \cite{62BK, 87Beatty}. Once a fully nonlinear elastic energy functional is available, expansion to third order in the
Green-Lagrange strain tensor $\B E$ and a direct comparison to equation (\ref{U_E}) yields $\{\beta_1, \beta_2, \beta_3\}$.

In Table \ref{tableII} we present the second order elastic constants for several materials (a glassy polymer, two metals and a soft material) using available experimental data or an expansion of a known fully nonlinear strain energy functional\footnote{The experimental papers reported the values of the Murnaghan coefficient's \cite{51Murnaghan}, which were transformed into $\{\beta_1, \beta_2, \beta_3\}$ using equation (\ref{Murnaghan}).}.
\begin{table}[here]
\caption{\label{tableII} The second order elastic constants $\{\beta_1, \beta_2, \beta_3\}$, in units of the shear modulus $\mu$, for various materials. The data for Polystyrene \cite{53HK}, copper and aluminium \cite{Crecraft} were obtained from acoustoelastic measurements. The constants for foam rubber were calculated using the fully nonlinear Blatz-Ko energy functional \cite{62BK, 87Beatty}.}
\footnotesize\rm
\begin{tabular*}{\textwidth}{@{}l*{15}{@{\extracolsep{0pt plus12pt}}l}}
\br
Material & $\beta_1/\mu$ & $\beta_2/\mu$ & $\beta_3/\mu$\\
\mr
\verb"Polystyrene"&-2.56&-6.01&-2.41\\
\verb"Copper"&5.17&-3.73 &-2.91&\\
\verb"Aluminium"&2.29 &-8.76 &-3.32\\
\verb"Foam rubber"&1/6 & -1&-8/3\\
\br
\end{tabular*}
\end{table}
Table \ref{tableII} reveals that the second order elastic constants have no definite sign and that they can be significantly larger than the linear elastic ones. While for (soft) foam rubber the second order elastic constants are of the order of the shear modulus, for ``hard'' solids they can be nearly an order of magnitude larger. Once $\{\beta_1, \beta_2, \beta_3\}$ are determined, the weakly nonlinear theory of dynamic fracture provides quantitative predictions that can be tested experimentally as described in section \ref{comparison}.

The second order elastic constants for an incompressible neo-Hookean material under plane-stress conditions will be discussed below  (cf. equation (\ref{constants_NH})).
\subsection{Properties of the weakly nonlinear solution}

Some properties of the weakly nonlinear solution are highlighted below. These properties represent non-trivial extensions of the linear solutions and, as we will show in section \ref{comparison}, can be tested experimentally.

%\subsubsection{Crack tip profiles in the weakly nonlinear theory}
%\label{CTOD}
\begin{itemize}
\item {\em Crack tip profiles in the weakly nonlinear theory} As was discussed above, an important and experimentally accessible property, is the crack tip opening profile (the so-called ``crack tip opening displacement'' -- CTOD). The weakly nonlinear theory predicts corrections to the parabolic CTOD predicted by LEFM, cf. equation (\ref{parabola}). Using equation (\ref{solution}), we obtain:
\begin{eqnarray}
\hspace{-2cm}u^{(2)}_x(r,\pm\pi) &=& \frac{K_I^2}{32\pi\mu^2}\left[(A+\alpha_s B)\log(r)+\sum_{n} c_n(v) \cos{(n\pi)}   \right]\equiv \chi_2 \log{(r)} + \chi_3 \ ,\nonumber\\
\label{opening_I}
\hspace{-2cm}u^{(2)}_y(r,\pm\pi) &=& \frac{\mp K_I^2(A+\alpha_d B)\pi}{32\pi\mu^2}\equiv \pm \chi_4 \ ,
\end{eqnarray}
which immediately implies
\begin{eqnarray}
\label{Tip_2nd}
\hspace{-2cm} \phi_x(r,\pm\pi) = -\chi_1\left(\phi_y(r,\pm\pi)\mp \chi_4 \right)^2 + \chi_2 \log{\left[\chi_1\left(\phi_y(r,\pm\pi)\mp \chi_4 \right)^2 \right]} + \chi_3 \ .
\end{eqnarray}
The latter is valid in the weakly nonlinear region and describes a parabolic form corrected by a logarithmic stretch. Note that by setting $\chi_2\!=\!\chi_3\!=\!\chi_4\!=\!0$, equation (\ref{parabola}) is recovered.

%\subsubsection{The sub-leading weakly nonlinear solution as a leading effect}
%\label{subleading}
\item
{\em The sub-leading weakly nonlinear solution as a leading effect} The second order displacement fields in equation (\ref{solution}) are the leading order corrections to the first order fields in equations (\ref{u_1st_I}) when a crack's tip is approached from the linear elastic region. There might exist situations in which these sub-leading weakly nonlinear contributions become the {\em dominant} ones, when the first order terms happen to vanish or become small for some physical reason. One such example was discussed in \cite{Harpaz.2012}, where it was shown that during mode II (shear) crack propagation, weakly nonlinear contributions are of tensile nature and hence may result in crack tip opening and tensile stresses ahead of the tip. This is a leading effect as these quantities are identically zero for the Mode II problem in LEFM. This effect may be relevant for problems like frictional sliding and the super-shear transition in mode II propagation \cite{Rosakis1999,Ben-David.10}.

Another example was briefly discussed in \cite{Livne.08, Bouchbinder.08a}, where it was noted that the tensile strain component $\pa_y u_y^{(1)}$ changes sign at a finite velocity, say $v_0$, ahead of a mode I crack's tip. Using equation (\ref{u_1st_I}) we obtain
\begin{equation}
\label{vanish_yy}
\pa_y u_y^{(1)}(r,\varphi\!=\!0) = \frac{K_I}{\mu \sqrt{2\pi r}} \frac{2\alpha_d\alpha_s-\alpha_d^2(1+\alpha_s^2)}{D(v)} \ ,
\end{equation}
which can be easily shown to change sign from positive to negative at
%$v_0\!=\!\left(\sqrt{c_d^2+8c_s^2} - c_d \right)/2$.
\begin{equation}
\label{v_0}
v_0 = \frac{1}{2} \left(\sqrt{c_d^2+8c_s^2} - c_d \right) \ .
\end{equation}
For example, for $c_d\!=\!2c_s$ we obtain $v_0\!=\!0.73c_s$, cf. figure \ref{strain_comparison}c \cite{Livne.08}.
For $v\!>\!v_0$, LEFM predicts that $\pa_y u_y^{(1)}(r,\varphi\!=\!0)$ is negative. This implies that, as the crack tip is approached, material points straddling $y \!=\!0$ come closer to one another instead of becoming increasingly separated as needed to precipitate fracture. This prediction of LEFM is not always appreciated. It was shown in \cite{Livne.08} to be in contrast with experimental measurements in which $\pa_y u_y^{(1)}(r,\varphi\!=\!0)$ is always positive for $v\!>\!v_0$. As we will show in figure \ref{strain_comparison}c,
the positive separation results from the dominant contribution provided by the weakly nonlinear theory; the second order contribution $\pa_y u_y^{(2)}(r,\varphi\!=\!0)$ is positive and dominates $\pa_y u_y^{(1)}(r,\varphi\!=\!0)$ in this velocity range.

%\subsubsection{The length scale associated with the weakly nonlinear theory}
%\label{lengthscale}
\item {\em The length scale associated with the weakly nonlinear theory}
One of the most important aspects of the weakly nonlinear theory of dynamic fracture is that it introduces a new length scale into the fracture problem. This length scale will be shown below to be of prime importance for understanding (at least) one dynamic crack tip instability, and might hold the key for cracking other puzzles in fracture dynamics. The new length scale, which we denote by $\ell_{nl}(v)$, represents the scale in which the nonlinear elastic contribution to the mechanical fields become comparable to the linear elastic asymptotic contributions near the tip of a crack. Physically, it describes the scale at which LEFM breaks down. The subscript $nl$ highlights the fact that the origin of this length scale is near tip nonlinearities.

To understand the properties of the length scale $\ell_{nl}(v)$, we expand the deformation gradient as $\B H \!\simeq\! \nabla_{\B x} \B u^{(1)} \!+\!  \nabla_{\B x} \B u^{(2)} \!\equiv\! \B H^{(1)} \!+\! \B H^{(2)}$ near the tip of a crack. $\ell_{nl}(v)$ emerges as a result of the different $r$-dependencies of $\B H^{(1)}$ and $\B H^{(2)}$ near the tip of a crack. According to equations (\ref{u_1st_I}), $\B H^{(1)}$ takes the following form
\begin{equation}
\B H^{(1)} = \B h^{(1)}(v/c_s,\lambda/\mu) \frac{K_I}{\mu\sqrt{r}} \ ,
\end{equation}
where $\B h^{(1)}(v/c_s,\lambda/\mu)$ is a calculable dimensionless tensorial function of the crack propagation speed $v/c_s$ and the ratio of the first order elastic constants $\lambda/\mu$.
According to equations (\ref{solution}), $\B H^{(2)}$ takes the following form
\begin{equation}
\B H^{(2)} = \B h^{(2)}(v/c_s,\lambda/\mu, \beta_1/\mu, \beta_2/\mu, \beta_3/\mu) \frac{K_I^2}{\mu^2\,r} \ ,
\end{equation}
where $\B h^{(2)}(v/c_s,\lambda/\mu, \beta_1/\mu, \beta_2/\mu, \beta_3/\mu)$ is a complicated, yet calculable, dimensionless tensorial function of the crack propagation speed $v/c_s$, the ratio of the first order elastic constants $\lambda/\mu$ and the dimensionless second order elastic constants $\{\beta_1/\mu, \beta_2/\mu, \beta_3/\mu \}$. $\ell_{nl}(v)$ can be estimated according to
\begin{equation}
\label{ell_nl}
|\B H^{(2)}|_{r=\ell_{nl}} \simeq |\B H^{(1)}|_{r=\ell_{nl}} \ ,
\end{equation}
which implies that {\em scaling-wise} it is given by
\begin{equation}
\label{ell_nl_1}
\ell_{nl} \sim \frac{K^2_I}{\mu^2} \sim \frac{\Gamma}{\mu} \ ,
\end{equation}
where we used the fact that the fracture energy $\Gamma$ is proportional to $K^2_I/\mu$ \cite{98Fre,Broberg.99}, (cf. equation (\ref{G(v)})). It is important to note that $\ell_{nl}$ is a {\em dynamic} length scale that evolves with the crack propagation speed $v$ due to both
the fracture energy $\Gamma(v)$ and the $v$-dependence of the pre-factor of equation (\ref{ell_nl_1}). The pre-factor, itself, is a nontrivial function of both $v$ and the first and second order elastic coefficients and is not necessarily of order unity.
Incorporating $\ell_{nl}$ into the theory of fracture, as will be discussed below, offers novel insight into the failure dynamics of solids and is one of the main take-home messages of this review.
\end{itemize}
\section{Comparing the theory to direct experimental measurements}
\label{comparison}

\begin{figure}[ht]
\includegraphics[width=0.95\columnwidth,clip=true,keepaspectratio=true]{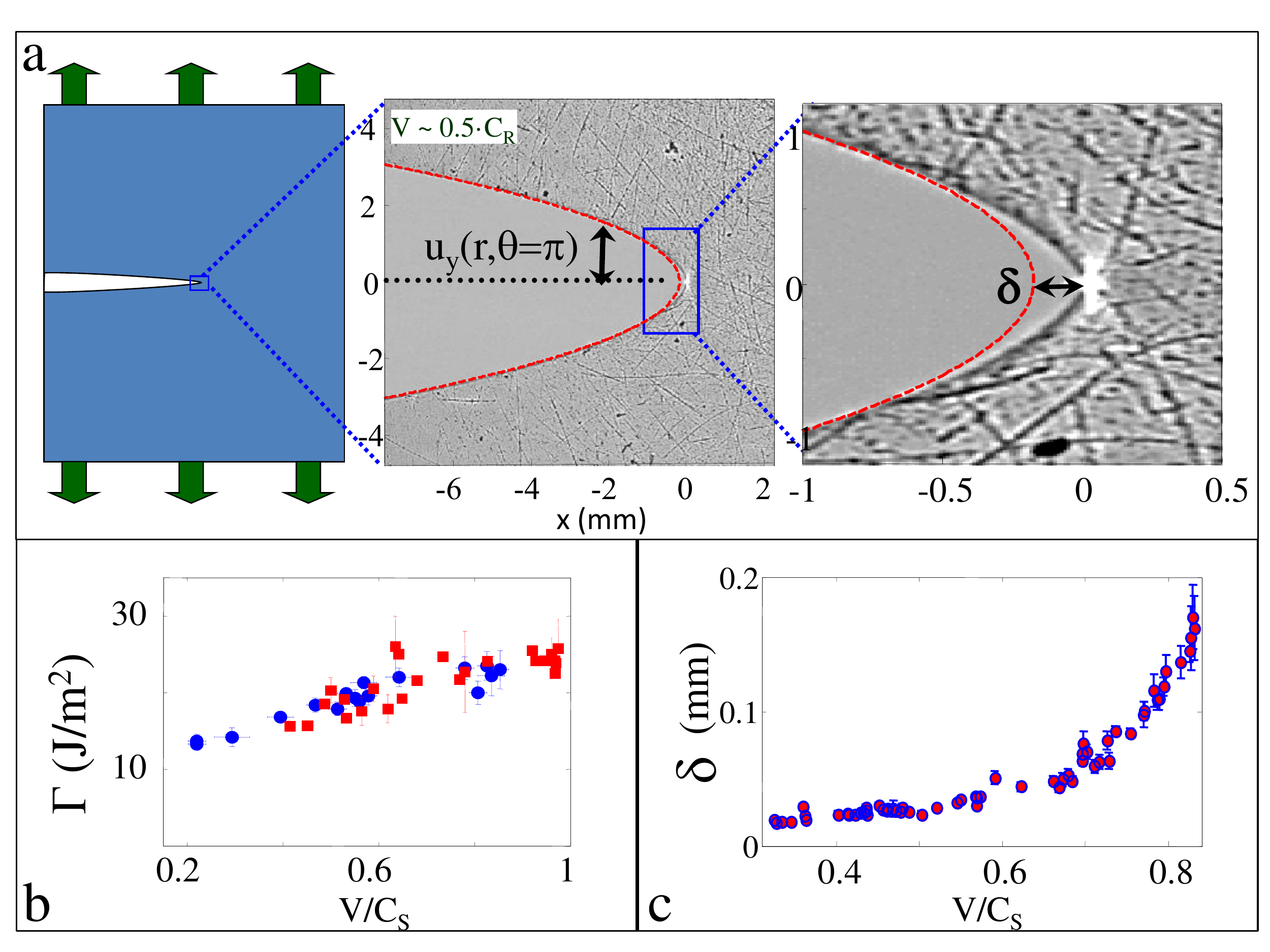}
\caption{In the close vicinity of the crack tip the pure parabolic form of the crack tip opening displacement (CTOD) predicted by LEFM  breaks down at a scale, $\delta$. (a) The crack tip opening at different scales: system scale (left) millimeter scale (center) sub-millimeter scale (right). The parabolic form predicted by equation \ref{parabola} (dashed line) fails to describe the near-tip region at scales below $\delta$, the distance from the real to the predicted tip of the crack.  (b) The fracture energy, $\Gamma$, obtained both from the crack tip curvature (equation (\ref{u_1st_I}) and using equation (\ref{G(v)}) (circles) and from steady state measurements in a strip (squares) (see \cite{Goldman.2010}). (c) $\delta(v)$ for the same velocity range. These results suggest that $\delta(v)$ is {\em not} a dissipative scale, as $\Gamma(v)$ is fairly constant with $v$ while $\delta(v)$ increases by nearly over an order of magnitude over the same velocity range.}  \label{NLintro}
\end{figure}

We now wish to use the soft gels discussed above to directly test the predictions of the weakly nonlinear theory of dynamic fracture. To accomplish this, we must first determine the second order elastic constants $\{\beta_1, \beta_2, \beta_3\}$ for polyacrylamide gels. The neo-Hookean constitutive law provides a good description of the nonlinear elastic behavior of many compliant materials \cite{75Tre}. This constitutive law is a natural extension of Hooke's law to finite deformations \cite{48R} and is the simplest description of rubber-like behavior based on the Gaussian chain statistical model of entropic elasticity \cite{75Tre}. The gels used in the experiments described below are incompressible and deform under plane-stress conditions (thin samples). Under these conditions the neo-Hookean energy functional takes the form \cite{83KS}
\begin{equation}
\label{NH} U(\B F)= \frac{\mu}{2}\left[tr(\B F^T\B F)+(\det\B F)^{-2}-3\right] \ ,
\end{equation}
where $\B F$ is the 2D deformation gradient tensor. $\det\B F$ appears in the elastic energy functional due to the incompressibility condition. Equation (\ref{NH}) conforms with the general expression in equation (\ref{U_E}) if one identifies $\lambda\!=\!2\mu$ and
\begin{eqnarray}
\label{constants_NH}
 \beta_1 = -4\mu/3,\quad \beta_2=0,\quad\beta_3 = -8\mu/3 \ .
\end{eqnarray}
With these values of the $\beta_i$'s in hand, we can now compare the general weakly nonlinear theory directly to experimental measurements. Let us first consider the measurements of the CTOD in the near vicinity of the crack tip, as presented in figure \ref{NLintro}. In figure \ref{instab_figure}b we saw that the crack tip can be described by a parabolic form, as predicted by equation (\ref{parabola}). As shown in equation (\ref{u_1st_I}), the crack tip curvature is wholly determined by $K_I$ and can be used to directly measure both $K_I$ and the fracture energy $\Gamma(v)$ (via  equation (\ref{G(v)})). Such measurements of $\Gamma(v)$ are not simply an exercise in curve fitting, but, as shown in figure \ref{NLintro}b, are in excellent agreement with independent measurements of $\Gamma(v)$ performed using a strip configuration \cite{Goldman.2010} (cf. equation (\ref{marder})) .

\begin{figure}[ht]
\includegraphics[width=0.95\columnwidth,clip=true,keepaspectratio=true]{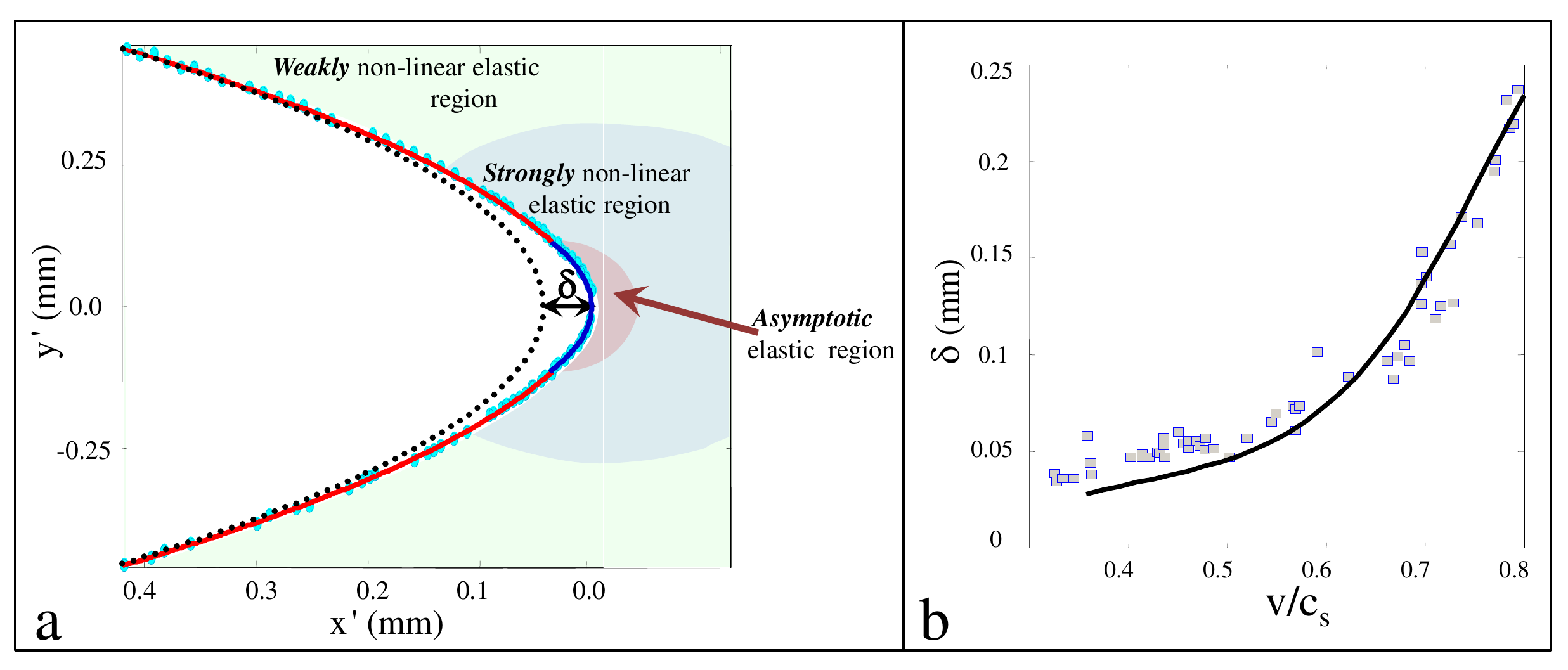}
\caption{The nonlinear elastic contributions to the CTOD. (a) A comparison of measurements (circles) at $v\sim 0.7c_R$ of the near-tip CTOD with predictions of LEFM (dotted line), the weakly non-linear theory (equation (\ref{Tip_2nd})) (red line) and theoretical predictions for a neo-Hookean material (blue line) in the nonlinear ``asymptotic'' region, where strains are significantly greater than unity (see \cite{Livne.2010} for more details). Nonlinear elastic contributions to the CTOD provide a nearly perfect description of the form of a crack's tip, as long as crack instabilities are suppressed and a single crack is propagating. (b) The same nonlinear elastic theoretical analysis (solid line) provides an excellent quantitative description of measured values (squares) of $\delta(v)$ \cite{Livne.2010}.}  \label{tip_comparison}
\end{figure}

Let us now look a bit more closely at the close vicinity of the crack tip. We find, as shown in the right panel of figure \ref{NLintro}a, that the CTOD predicted by LEFM fails to describe the crack tip within a region $\delta$, defined as the distance between the real crack tip and the LEFM prediction. As LEFM predictions perfectly describe both the CTOD at intermediate scales (e.g. center panel of figure \ref{NLintro}a) and $\Gamma(v)$ (figure \ref{NLintro}b), the existence of $\delta$ is not an ``artifact" of a ``poor" fit to the data. In fact, figure \ref{NLintro}c demonstrates that $\delta(v)$ is a systematically increasing function of $v$, whose value increases by over an order of magnitude as $v$ doubles. Comparison of the steep increase in $\delta(v)$ with the mildly varying $\Gamma(v)$ over the same velocity range, suggests that the increase in $\delta$ is {\em not} due to an increased dissipation with $v$, but is, instead, an effect of elastic nonlinearity. A large part of this effect, as described in weakly nonlinear solution, equation (\ref{Tip_2nd}), is  a ``stretching" of the $x$ axis due to the new logarithmic term resulting from weak nonlinear elasticity. This is demonstrated in figure \ref{tip_comparison}, where we present a detailed comparison of LEFM predictions (dotted line) and those of the weakly nonlinear theory (red line) given by  equation (\ref{Tip_2nd}). For completeness, we show the concatenation of the weakly nonlinear theory to predictions of an asymptotic theory \cite{Livne.2010,Knowles.83} accounting for strong elastic nonlinearity (at scales within $100\mu m$ from the tip) in neo-Hookean materials that are encountered at very large ($\gg 1$) strains. The theoretical comparison described in  figure \ref{tip_comparison}a was obtained from first principles using {\em no} adjustable parameters, as the second order elastic constants for neo-Hookean materials are known (cf. equation (\ref{constants_NH})).

The same analysis that yielded figure \ref{tip_comparison}a for a particular crack velocity was performed for the range of $v$ studied in \cite{Livne.2010}. This analysis resulted in the excellent quantitative agreement with the measured values of $\delta(v)$ presented in figure \ref{tip_comparison}b. Figure \ref{tip_comparison} therefore demonstrates that simply accounting for the nonlinear elasticity of the material surrounding a crack's tip provides us with a complete quantitative description of the CTOD of highly dynamic cracks.

\begin{figure}[ht]
\includegraphics[width=0.95\columnwidth,clip=true,keepaspectratio=true]{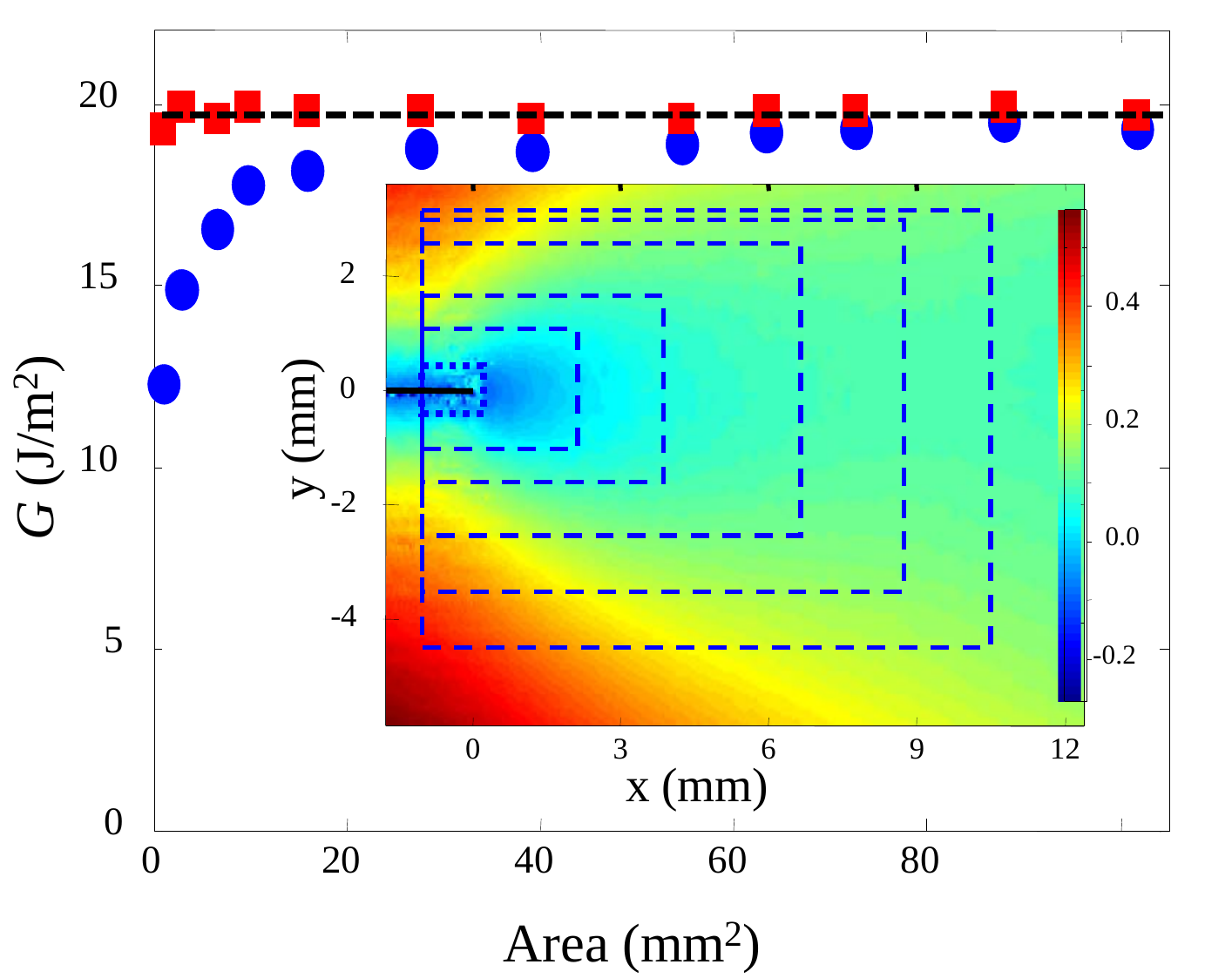}
\caption{The energy flux $G$ through different contours using the neo-Hookean energy functional in equation (\ref{NH}) (red squares) and its linear elastic approximation (blue circles). (Inset) The measured displacement
field $u_x$ of a crack propagating at $v\!\simeq\!0.7c_s$ (color bar in millimeters). Dashed blue rectangles mark every second contour used for calculating $G$. The value of $G$ corresponds to the independently measured value of $\Gamma(v\!\simeq\!0.7c_s)$ (dashed black line). The results are presented in the reference (undeformed) configuration, where the crack is denoted by the black line with its tip at the origin. Adapted from \cite{Livne.2010}.}  \label{J_integral}
\end{figure}

At what scale do elastic nonlinearities and dissipation take place in the neo-Hookean materials considered? Figure \ref{J_integral} describes direct measurements of the integrated energy flowing into contours of decreasing size that surround the tip of a crack propagating at $v\!\simeq\! 0.7c_s$ \cite{Livne.2010}. The experimentally measured displacement fields enable us to directly calculate the energy flux through any contour ${\C C}$ {\em assuming} a specific elastic energy functional $U$, using the J-integral \cite{Rice.68, 98Fre}
\begin{equation}
\label{Jint}
J=\int_{\C C} \left[\left(U+\case{1}{2}\rho_0 \pa_t u_i \pa_t u_i\right)v n_x+ s_{ij} n_j\pa_t u_i\right]d{\C C}\ ,
\end{equation}
where $\B n$ here is an outward unit vector on ${\C C}$. The energy release rate is given by $G\!=\!J/v$ \cite{Rice.68}. This calculation puts a stringent constraint on $U$ since only the physically correct one will make the result path (contour) independent, as must be the case as long as the path does not go through a dissipative region and the crack propagates at a steady velocity. The squares in figure \ref{J_integral} correspond to $G$ for different contours obtained using the elastic energy function of equation (\ref{NH}), while the circles represent the corresponding calculations using the linear elastic approximation of equation (\ref{U_varepsilon}). To the best of our knowledge, this is the first ever direct experimental estimate of the J-integral for a moving crack. This figure demonstrates three important things: (i) The elastic energy function of equation (\ref{NH}) properly describes the constitutive behaviour of this material down to the smallest scales near the tip of the crack since $G$ is indeed shown to be path (contour) independent with a value exactly equal to the measured fracture energy $\Gamma(v)$ at this propagation velocity. (ii) The progressive breakdown of LEFM as the crack tip is approached highlights the existence of the length scale, $\ell_{nl}$, associated with near tip elastic nonlinearities. At large scales (8-10mm) the deformation is small and nonlinearities in $U$ make no difference (the circles overlap the squares). As the tip is approached (i.e. smaller and smaller contours are used), the deformation becomes progressively nonlinear and the LEFM predictions deviate significantly from the nonlinear ones (reaching about 50\% deviation for the smallest contour of typical length of a few hundred $\mu$m. (iii) Down to the smallest contour, no dissipation is observed ($G$ in the nonlinear calculation is still constant, cf. the squares). Hence, an upper bound for the dissipative scale in this material is a few hundred $\mu$m. Comparison to the nonlinear ``asymptotic'' theory in further reduces the upper bound of the size of the dissipative region to within $\sim 20\mu$m from the crack tip \cite{Livne.2010}.

\begin{figure}[ht]
\includegraphics[width=0.95\columnwidth,clip=true,keepaspectratio=true]{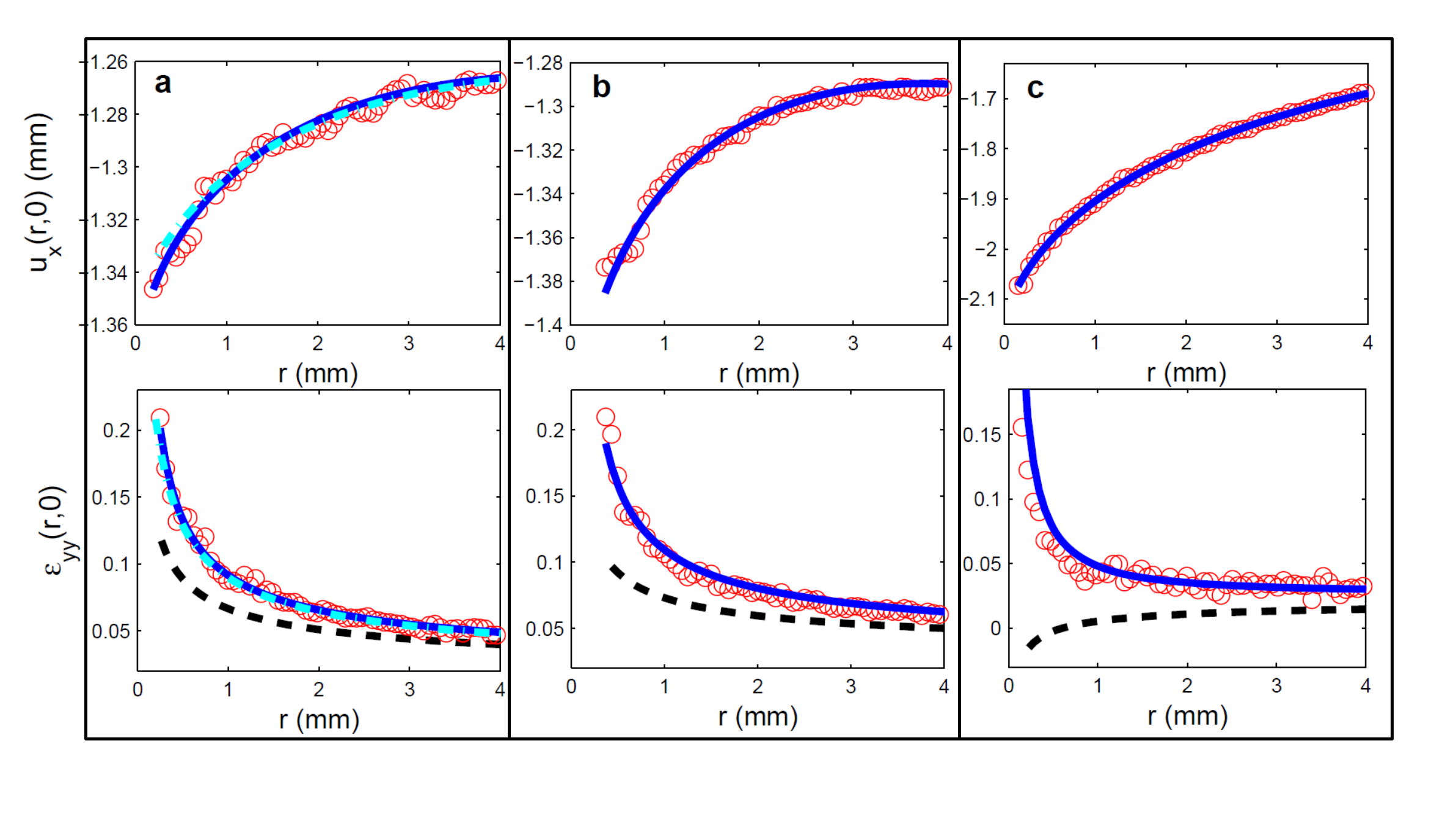}
\caption {Top: Measured $u_x(r,0)$ (circles) fitted to
the x component of equation (\ref{solution}) (solid line) for (a) $v\!=\!0.20c_s$ with $K_I\!=\!1070$Pa$\sqrt{m}$, $T\!=\!-3150$Pa and $B\!=\!18$. (b) $v\!=\!0.53c_s$ with $K_I\!=\!1250$Pa$\sqrt{m}$, $T\!=\!-6200$Pa and $B\!=\!7.3$ and (c) $v\!=\!0.78c_s$ with $K_I\!=980$Pa$\sqrt{m}$, $T\!=\!-6900$Pa and $B\!=\!26$ Bottom: corresponding measurements of $\varepsilon_{yy}(r,0)\!=\!\pa_y u_y(r,0)$ (circles) compared to the  weakly nonlinear solution (solid lines) (cf. equation (\ref{expansion})) where $K_I$, $T$ and $B$ are taken from the fits of
$u_x(r,0)$ and $u_y(r,\pi)$. (dashed lines) LEFM predictions (analysis as in \cite{Livne.08}) were added for comparison. In the left panels we added (dashed-dotted lines) the weakly nonlinear solution {\em where the nonlinear parameters are theoretically calculated} using $K_I\!=\!1040$Pa$\sqrt{m}$ and $T\!=\!-2800$Pa. The results are nearly indistinguishable from those obtained by taking the nonlinear parameter $B$ (in equation (\ref{A_B_bc})) as a free parameter (solid lines). Adapted from \cite{Livne.2010b}.}  \label{strain_comparison}
\end{figure}

We now wish to perform a direct comparison with the strain fields that are measured as close to the crack tip as possible. As we saw in equation (\ref{1/r}), the nonlinear corrections to the strains have a stronger singularity ($\sim \!1/r$) than the $\sim\!1/\sqrt{r}$ singularity predicted by LEFM. In addition, at high velocities ($\sim \!0.7c_R$) the nonlinear contribution to the strain becomes the {\em dominant} one  (cf. equation (\ref{vanish_yy})). In figure \ref{strain_comparison} we compare direct measurements of both $u_x(r,0)$ and $\varepsilon_{yy}(r,0)\!=\!\pa_y u_y(r,0)$ to predictions of the weakly nonlinear theory (line) and LEFM (dotted line).
The explicit form of the weakly nonlinear solution in equation (\ref{solution}) for a neo-Hookean material contains only two parameters ($K_I$, $T$) that cannot be determined from the asymptotic solution and therefore must be extracted from the experimental data. $K_I$ was determined by fitting the far-field parabolic form of the CTOD to equation (\ref{opening_I}). The value of $T$ for each $v$ is unknown and was left as a free parameter \cite{Bouchbinder.08a}. We then follow \cite{Bouchbinder.08a} and use equation (\ref{solution}) with the nonlinear parameter $B$ in equation (\ref{A_B_bc}) as a free fitting parameter to compare the theory to the measured experimental strains. For $v \!=\!0.20 c_s$ we also include the predicted curves using the {\em theoretically calculated} value of $B$. In figure \ref{strain_comparison} we present the resulting comparison for $u_x(r,0)$ and $\varepsilon_{yy}(r,0)$ with $v/c_s\!=\!0.20$, $0.53$ and $0.78$.

It is evident from figure \ref{strain_comparison} that the agreement with the experimental data is excellent. For the low velocity curve there is essentially no difference between the results corresponding to the calculated and fitted values of $B$. The results clearly demonstrate the effect of the predicted $1/r$ singular terms near the crack tip. In particular, the highest velocity ($v\!=\!0.78c_s$) is larger than $v_0\!=\!0.73c_s$ calculated in equation (\ref{v_0}), for which LEFM predicts that $\varepsilon_{yy}(r,0)$ becomes {\em negative} (cf. the dashed line in panel c). We see that the second order theory already avoids this intuitive conundrum for $v\!>\!v_0$.  Thus, the second order nonlinear solution (solid line) both induces a qualitative change in the character of the strain and, moreover, yields excellent quantitative agreement to the measurements. This favorable comparison demonstrates that that the contributions of elastic nonlinearities are both important and generally unavoidable. This is especially true as high crack velocities are reached and the size of the nonlinear region, as evidenced by $\delta$, becomes large.

\section{Cracking a dynamic crack instability}

In the previous section we demonstrated that the weakly nonlinear theory is a natural and a necessary extension of LEFM, providing an excellent quantitative description of the near-tip fields surrounding a propagating crack. These results provide a new and comprehensive picture of how remotely applied forces drive material failure in the most fundamental of fracture states: straight, rapidly moving cracks. In particular, these results reveal the beauty of how the hierarchy of linear and nonlinear elastic zones couple to transport energy from the macroscopic scales inherent in the external loading to the increasingly singular regions surrounding a crack's tip -  before being dissipated at still smaller scales. On the other hand, as long as a single crack retains its path stability the theory provides nothing new with regard to a crack's dynamics; energy-balance essentially governs the overall behavior of a rapidly moving ``simple" crack.

Were dynamic cracks always simple cracks, we would now have a complete theory of fracture up to the dissipative scale. As we have shown in section \ref{instabilities}, however, simple cracks become unstable in an number of ways. Furthermore, in the case of the oscillatory instability, a new length scale (the oscillations wavelength) was born, whose origin had nothing to do with all typical external scales in the system. We have shown that the weakly nonlinear theory provides a new and intrinsic length scale, $\ell_{nl}$ (cf. equation (\ref{ell_nl_1})). This scale is determined dynamically; it represents the crossover length where nonlinear elastic effects become significant.  Experimentally, $\ell_{nl}$ is simply related to the scale $\delta(v)$, which describes the ``extra" length that a crack tip receives due to the nonlinear elastic contributions.

In a simple crack $\ell_{nl}$ is a ``passive'' quantity. Although formed by the nonlinear elastic fields, it is simply carried along at the tip of the crack. In this section we explore the ramifications when this scale is no longer passive, becoming instead {\em dynamic} (or active). When this occurs, non-trivial interactions can take place between the linear fields that drive the non-linear region and the non-linear response, as represented by $\ell_{nl}$. We shall see that this active feedback between the linear and nonlinear regions provides the key towards understanding the oscillatory instability.

\subsection{A dynamic crack tip equation of motion and linear stability analysis}
\label{tip_eom}

Probably the first question one should raise in relation to dynamic fracture instabilities is why our understanding of them is much less developed in comparison to other, seemingly similar, instabilities in condensed matter physics and material science? For example, the micro-branching crack instability might appear somewhat similar to side-branching in dendritic crystal growth \cite{mullins1964stability}. The linear regime of the solidification instability was essentially explained by Mullins and Sekerka nearly fifty years ago \cite{mullins1964stability}, and that insight has been the basis for major advances in solidification theory and processing ever since \cite{asta2009solidification}. This progress has stemmed directly from the fact that the dynamical evolution of the solid-liquid interface is governed on a continuum scale by a well-defined free-boundary problem. In contrast, it is not yet clear how to rigorously formulate an analogous free-boundary problem for fracture, where material failure is localized to a small singular region near the crack tip. Consequently we have, as of yet, no comparable understanding of dynamic fracture instabilities. One crucial missing ingredient is an understanding of the time and length scales associated with the physics near the tip of a propagating crack. In this section we will describe a recent attempt to derive a new dynamic crack tip equation of motion, incorporating the existence of the length scale $\ell_{nl}(v)$ \cite{Bouchbinder.09b}.

%%%%%% FIGURE %%%%%%%%%%%%%%%%%%
\begin{figure}
\centering \epsfig{width=.6\textwidth,file=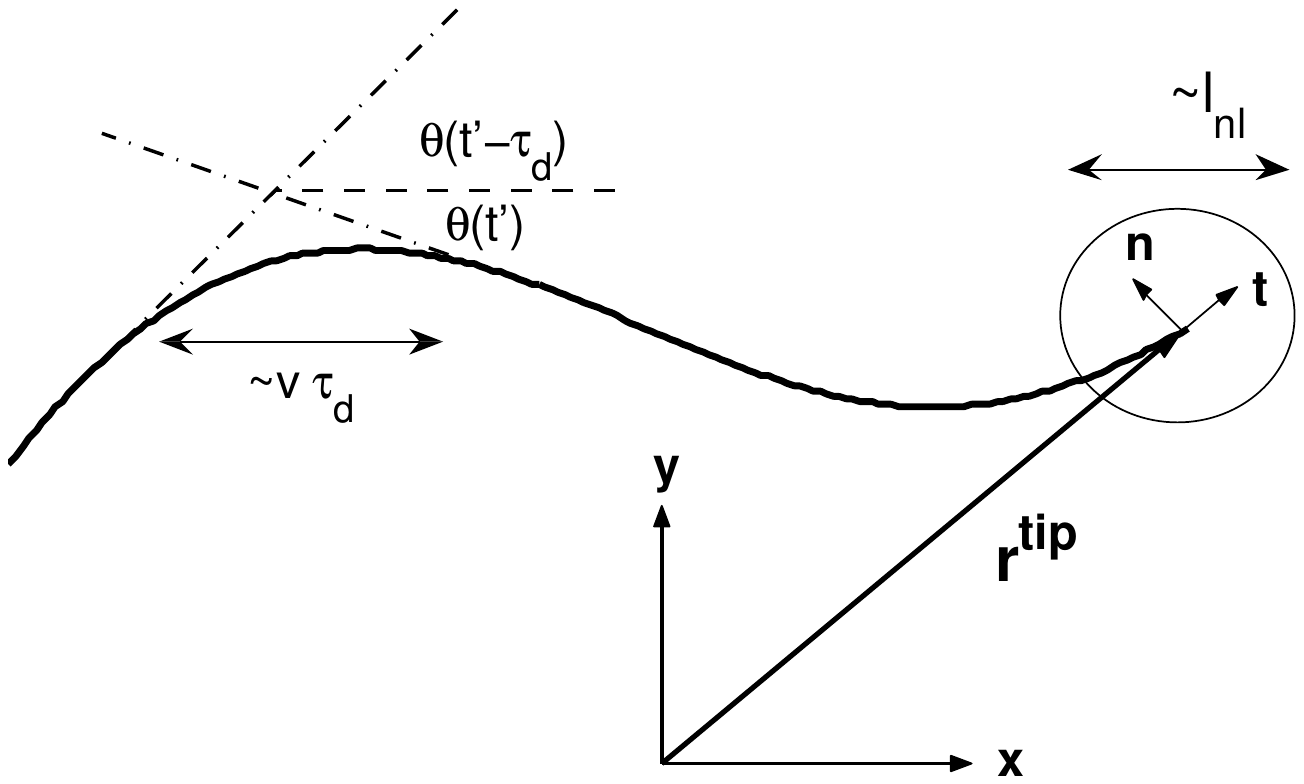} \caption{A crack with a small nonlinear zone of scale $\ell_{nl}$. The angle $\theta$ the crack makes with respect to the x-direction is shown at two times separated by a delay $\tau_d$.}\label{sketch}
\end{figure}
%%%%%%%%%%%%%%%%%%%%%%%%%%%%%%%%%%%

The first step in this derivation closely follows the reasoning of \cite{Hodgdon.93}. Consider a crack in a 2D body, whose path is described by $\B r^{tip}(t)$ and whose tip is surrounded by a small nonlinear zone of scale $\ell_{nl}$. We denote by $\B t$ and $\B n$ the tangent and normal unit vectors at the crack tip, respectively (see figure \ref{sketch}). As we consider an arbitrary crack path under general external loading conditions, the LEFM near tip fields are generically characterized by both $K_I$ (tension) and $K_{II}$ (shear) contributions. Consider then the discrete symmetry operation $R_n$ that transforms $\B n\! \rightarrow\! -\B n$. Under this symmetry operation, the relevant quantities of the asymptotic LEFM fields for both mode I and II fracture transform as follows:
(i) $K_I \!\rightarrow\! K_I$ (ii) $K_{II}\! \rightarrow \!-K_{II}$ (iii) $v\! \rightarrow \!v$.

Assuming material isotropy, one can write down the most general first order equations that are invariant under $R_n$.
The first equation is just a kinematic relation for the rate of crack growth
\begin{equation}
\label{translation}
\partial_t {\bf r}^{tip} =  v(K_I(t),K_{II}(t))~\!\B t \ .
\end{equation}
The second one describes crack tip rotation
\begin{equation}
\label{rotation}
\partial_t \B t \propto K_{II}(t) \B n \ ,
\end{equation}
where the proportionality coefficient is a true scalar.

The second step in the derivation amounts to estimating the proportionality coefficient in equation (\ref{rotation}) by dimensional considerations. The existence of a length scale $\ell_{nl}$ and a crack speed $v$ suggests a time scale $\ell_{nl}/v$. Equation (\ref{initiation}) allows us to define a typical stress intensity factor as $\bar{K}_c \!\sim\! \sqrt{E\Gamma}$. Together, these imply that equation (\ref{rotation}) can be written as
\begin{equation}
\label{rotation1}
\partial_t \B t \simeq - \frac{v}{\ell_{nl}} \frac{K_{II}(t)}{\bar{K}_c} \B n \ .
\end{equation}
The adopted scaling approach assumes that all other material-specific properties of the nonlinear zone appear as a pre-factor of order unity in equation (\ref{rotation1}).
Equation (\ref{rotation1}) can be rewritten in terms of the angle $\theta$ that the unit tangent $\B t$ makes with the
x-axis as \cite{Bouchbinder.07, Bouchbinder.03}
\begin{equation}
\label{rotation2} \partial_t \theta(t)\simeq -\frac{v}{\ell_{nl}} \frac{K_{II}(t)}{\bar{K}_c} \ .
\end{equation}

The third step in the derivation follows from the observation that $\theta(t)$ is defined at the crack tip, while $K_{II}(t)$ is defined a distance $\sim \ell_{nl}$ away from it. Therefore, causality implies that $K_{II}$ at time $t$ cannot be affected by the crack faces created in the time interval $[t\!-\!\tau_d,t]$, with
\begin{equation}
\label{tau}
\tau_d \sim \ell_{nl}/c_{nl} \ .
\end{equation}
Here $c_{nl}$ is the typical wave speed within the nonlinear zone, possibly of the order of the linear elastic wave speed $c_s$, but not necessarily so. We note that using a single delay time $\tau_d$ is certainly a simplification of more complicated dynamics, but this simplified scaling assumption is expected to capture the essence of the physics involved.

To formulate this idea precisely, we should express the {\em physical} $K_{II}$ at time $t$ in terms of the {\em mathematical} $\C K_{I}$ and $\C K_{II}$ at a {\em retarded time} $t\!-\!\tau_d$, taking into account the fact that the latter are defined with respect to a coordinate system rotated by $\theta(t\!-\!\tau_d)$, while the former with respect to a coordinates system rotated by $\theta(t)$, see figure \ref{sketch}. $\C K_{I}(t\!-\!\tau_d)$ and $\C K_{II}(t\!-\!\tau_d)$ are obtained from a pure LEFM problem with a crack path corresponding to $t\!-\!\tau_d$ and a simple consideration allows us to express $K_{II}(t)$ in terms of them. Equation (\ref{rotation2}), supplemented by the relation $K_{II}(t)\left[\C K_{I}(t\!-\!\tau_d), \C K_{II}(t\!-\!\tau_d) \right]$, constitutes the proposed dynamic equation of motion for the crack tip. It can be shown to reduce to the so-called ``principle of local symmetry'' -- stating that cracks propagate so as to annihilate $K_{II}$ \cite{Hakim.09, Goldstein.Salganik,Cotterell.Rice,Lazarus2001, Adda-Bedia.95, Pham.08, Corson.09, hakim2005crack} -- under quasi-static conditions. In this limit, the delay time $\tau_d$ is expected to play no important role. For propagation velocities of the order of the speed of information, however, new physical effects might emerge. It is important to stress that the derivation of equation (\ref{rotation2}) did not make any explicit reference to the {\em origin} of $\ell_{nl}$, which denotes the scale in which linear elasticity breaks down, either by nonlinear elasticity or by dissipative process. For example, in principle it could equally well apply to situations in which the breakdown of linear elasticity is dominated by plastic deformation. This remains to be verified experimentally in the future.

\begin{figure}
\centering \epsfig{width=.6\textwidth,file=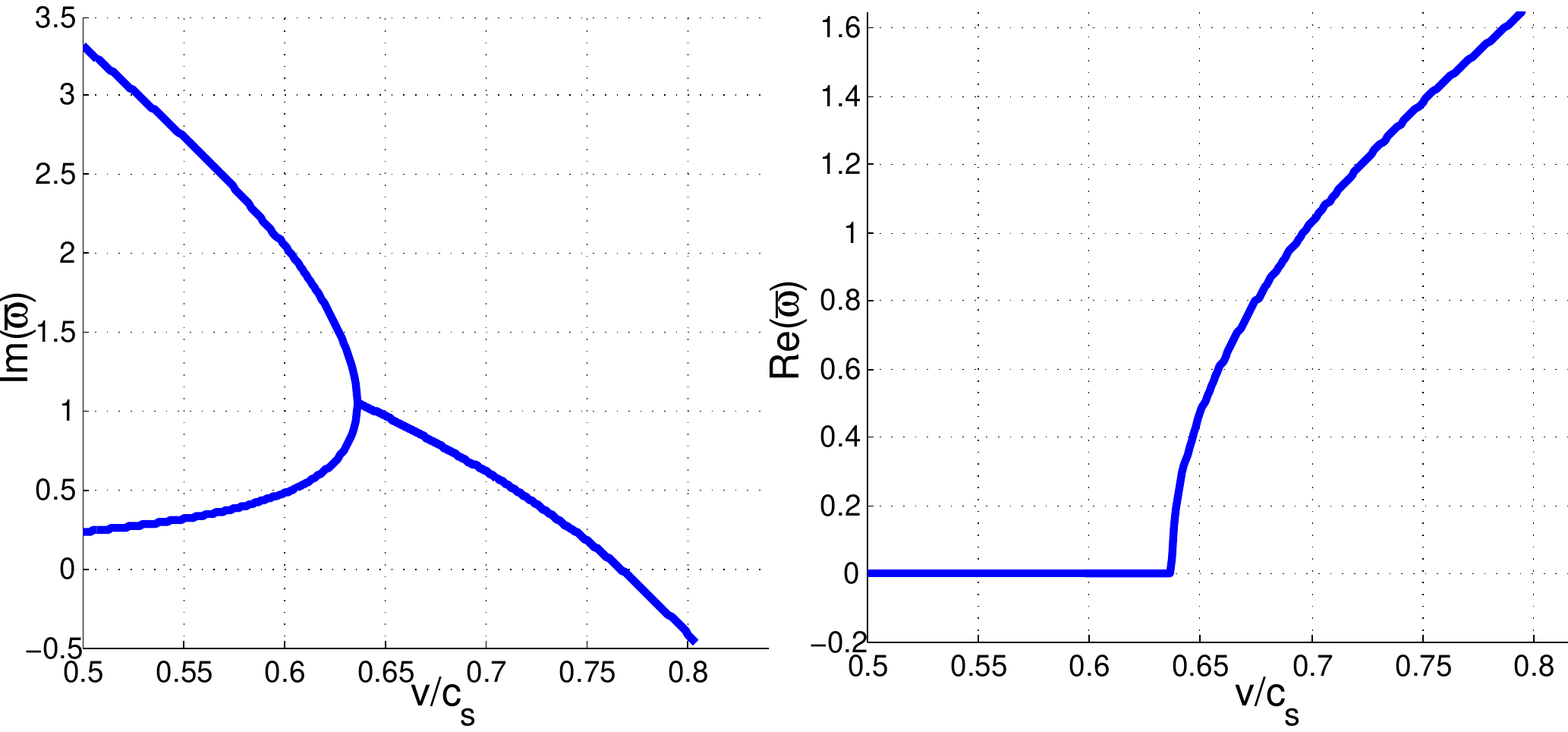} \caption{$\Im(\bar{\omega})$ (left) and $\Re(\bar{\omega})$ (right) as a function of $v/c_s$, where $\bar{\omega}\!\equiv\!\omega\tau_d$ (see \cite{Bouchbinder.09b} for more details). The existence of a linear oscillatory instability is predicted at $v_c\!\simeq\!0.77c_s$, for which $\Im(\bar{\omega})$ becomes negative with $\Re(\bar{\omega})\!\ne\!0$.} \label{instability}
\end{figure}

Equation (\ref{rotation2}) is a nonlinear integro-differential equation for $\theta(t)$. With such an equation at hand, one can mathematically pose the question of the linear stability of the crack's path. To address this, consider a straight crack propagating at a steady velocity $v$ under mode I symmetry conditions. Consider then a small perturbation of the straight path, characterized by an amplitude $a$ and a wavelength $\lambda$. In the limit $a/\lambda \!\ll\!1$, we can consider linear modes of the form $\theta(t)\!\simeq \!a e^{i \omega t}/\lambda$, with $\Re{(\omega)}\!=\! 2\pi v/\lambda$, and linearize equation (\ref{rotation2}) with respect to $a/\lambda$. The resulting equation for $\omega$ determines the linear stability of the crack, employing the Willis-Movchan linear perturbation formalism \cite{Willis_Mochvan.95, Willis.02, Bouchbinder.09b}. In particular, $\Im(\omega)\!<\!0$ implies stability as path perturbations decay in time, while $\Im(\omega)\!<\!0$ implies an instability as path perturbations are amplified. The real and imaginary parts of the complex angular frequency $\omega$, as a function of $v$, are shown in figure \ref{instability}. We observe that $\Im(\omega)$ changes sign from positive to negative at high velocity $v_c$ close to $c_s$. The critical velocity $v_c$ is only weakly dependent on the Poisson ratio (not shown). Moreover, $\Re(\bar{\omega})\!\ne\!0$ at this velocity. Together, these imply a high-velocity oscillatory instability with a wavelength $\lambda_{osc}$ that satisfies

\begin{equation}
\lambda_{osc} \sim \tau_d v_c \sim \ell_{nl} \ .
\end{equation}
This is an important prediction, suggesting the existence of a high-velocity oscillatory instability that is controlled by {\em intrinsic} time and length scales associated with the breakdown of LEFM near the tip of a crack.
The critical velocity $v_c$ is about 80\% of the shear wave-speed, but in light of the various approximations adopted, we do not intend to compare its exact value to experimental data.
% ******************deleted******************************
% Instead, we aim at testing the generic predications of the theory, namely the existence of an oscillatory instability at a normalized velocity $v_c/c_s$ close to unity and independent
%of the material properties, with a wavelength $\lambda_{osc}$ that scales with the size of the nonlinear zone $\ell_{nl}$ around the tip.
%***********************************************************

\subsection{Experimental test}

The theory described in the previous section predicts that, in the presence of a finite $\ell_{nl}$, causality implies that the singular LEFM fields lag behind the actual tip location with a delay time $\tau_d\!\propto\!\ell_{nl}$. This is linked to a high-velocity oscillatory instability with the following properties:
\begin{itemize}
\item  The scaled critical velocity for the onset of oscillations $v_c/c_s$ is close to unity and material independent.
\item  The oscillation wavelength $\lambda_{osc}$ is proportional to $\ell_{nl}$.
\end{itemize}

\begin{figure}
\centering \epsfig{width=1\textwidth,file=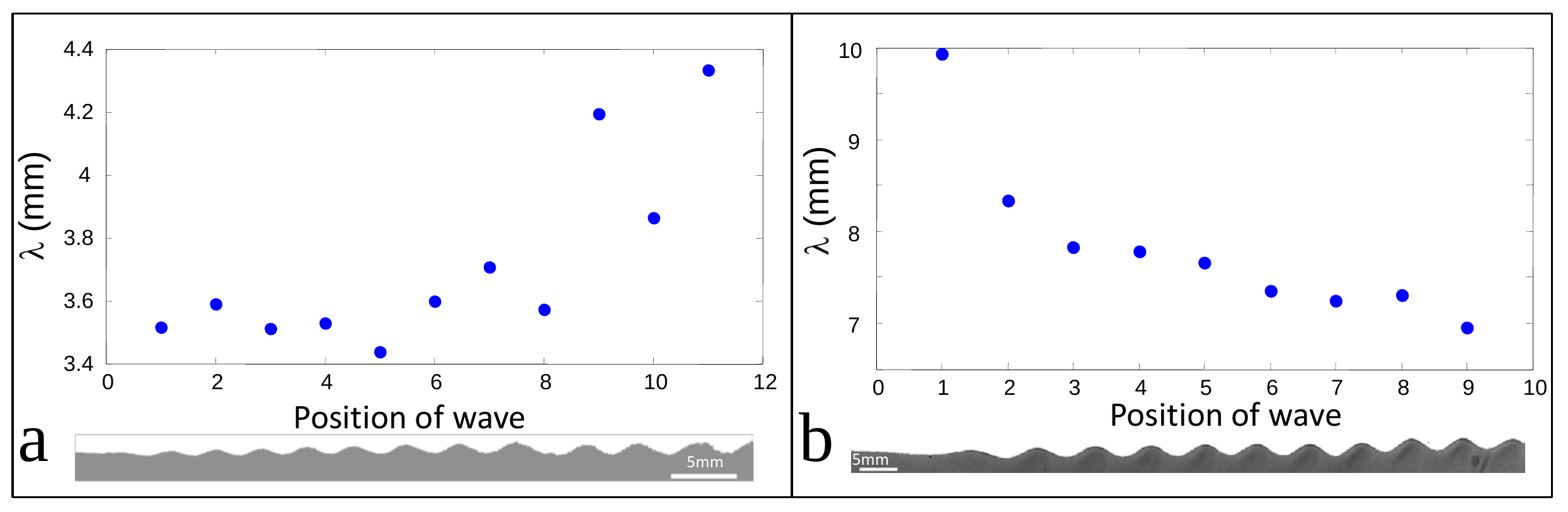} \caption{The oscillation wavelength, $\lambda_{osc}$ changes significantly as the instability develops. Shown are typical experiments depicting the wavelength evolution with their order of appearance for materials with (a) $\Gamma=39J/m^2$, $\mu=168kPa$ (b) $\Gamma=27J/m^2$, $\mu=36kPa$. (bottom) Typical photographs of the $xy$ profiles of fracture surfaces corresponding to the data sets.}\label{choose_wavelength}
\end{figure}

The best way to test these predictions is to excite the oscillatory instability in a variety of different materials, thereby controllably varying both $\mu$ and $\Gamma(v)$ (and hence $\ell_{nl}$, cf. equation (\ref{ell_nl_1})). These experiments were performed in \cite{Goldman.12} by preparing a variety of polyacrylamide gels, by varying the concentration of the gel components, and thereby changing both the elastic moduli and $\Gamma$ for each gel composition. In this way we constructed gels over a range of shear moduli $\mu$ ($33\!<\!\mu\!<\!187$\,kPa) and fracture energies $24\!<\!\Gamma(v_c)\!<\!60$\,J/m$^2$ at the critical velocity for the onset of the instability, $v_c$. The experiments were performed in thin sheets of dimensions ($x\times y\times z$) ($130\times 130\times 0.2$)mm and ($200\times 200\times 0.2$)mm, where as before $x$, $y$ and $z$ are, respectively, the propagation, loading and thickness directions. These dimensions were large enough so that the dynamics at $v_c$ occurred in an effectively infinite 2D medium. The sheet thickness was small enough ($160\!-\!220\mu$m) to suppress the micro-branching instability to enable single crack states to attain velocities to beyond $0.9c_s$ (as in \cite{Livne.07}). Once excited, the oscillatory wavelength, $\lambda_{osc}$, may change significantly as the instability evolves, as demonstrated in figure \ref{choose_wavelength}. In order to test the theoretically predicted initial wavelength, comparison with the theory was performed using only the {\em first} excited wavelength.

\begin{figure}
\centering \epsfig{width=1\textwidth,file=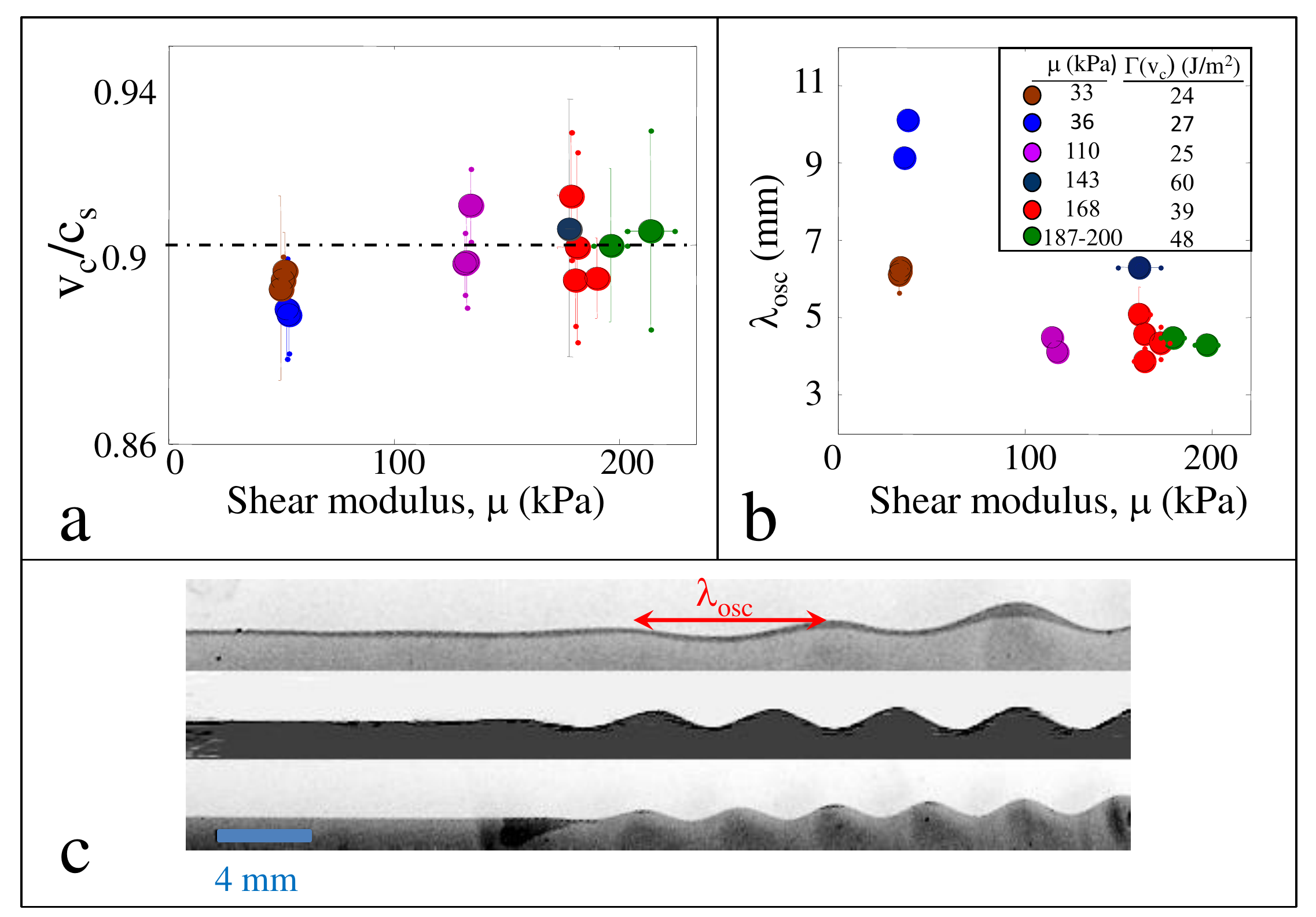} \caption{(a) The scaled critical velocity, at the onset of the instability $v_c\!\simeq\!0.9c_s$, is constant. $v_c$ is defined as the maximal velocity prior to the instability onset in each material. Symbol colors correspond to the legend of panel b (b) The oscillation wavelength, $\lambda_{osc}$ changes significantly with the material, as characterized by $\mu$. (c) Typical photographs of the $xy$ profiles of fracture surfaces at the onset of the oscillatory instability; from top to bottom: $\mu$= 36, 143, and 168 kPa. Figure adapted from \cite{Goldman.12}.}\label{1st_wavelength}
\end{figure}

In each of the experiments shown, experimental conditions were identical except for the material used. The first prediction of the theory is that the scaled critical velocity, $v_c/c_s$, at the onset of the oscillatory instability is of order unity and material independent. This is born out in figure \ref{1st_wavelength}a, which demonstrates that $v_c\approx 0.9c_s$ for all of the materials tested.
Figures \ref{1st_wavelength}b and c demonstrate that variation of the material, on the other hand, causes a wide variation of the of the oscillatory wavelength, $\lambda_{osc}$, with the elastic modulus of the material. We wish to compare $\lambda_{osc}$ to the non-linear scale, $\ell_{nl}$, predicted by the theory.

There are a number of ways to estimate or measure $\ell_{nl}$. One method is by means of equation (\ref{ell_nl_1}) which states that $\ell_{nl} \propto \Gamma/\mu$, though with a highly nontrivial pre-factor.  Using this loading configuration, $\Gamma(v)$ could be measured using equation (\ref{G(v)}), by determining $K_I(v_c)$ via the crack-tip curvature (cf. equation (\ref{opening_I})), when corrected for finite strain \cite{Goldman.12}.
$\Gamma(v_c)$ is determined by measuring the crack tip curvature at $v_c$ and utilizing the universal kinematic functions calculated in the framework of LEFM. These functions (see equation \ref{G(v)}) become singular as $v\rightarrow c_R$. As a result, the value of $\Gamma(v_c)$ can be quite sensitive to small uncertainties in $c_R$.

An additional method for estimating the scale of $\ell_{nl}$ is by directly measuring $\delta(v_c)$ in each material. $\delta(v_c)$ is a scale that incorporates all of the nonlinear elastic effects in the vicinity of the crack tip, and therefore must be proportional to $\ell_{nl}$.
This method has the distinct advantage of using a directly measurable quantity, with relatively small uncertainties.

\begin{figure}
\centering \epsfig{width=.8\textwidth,file=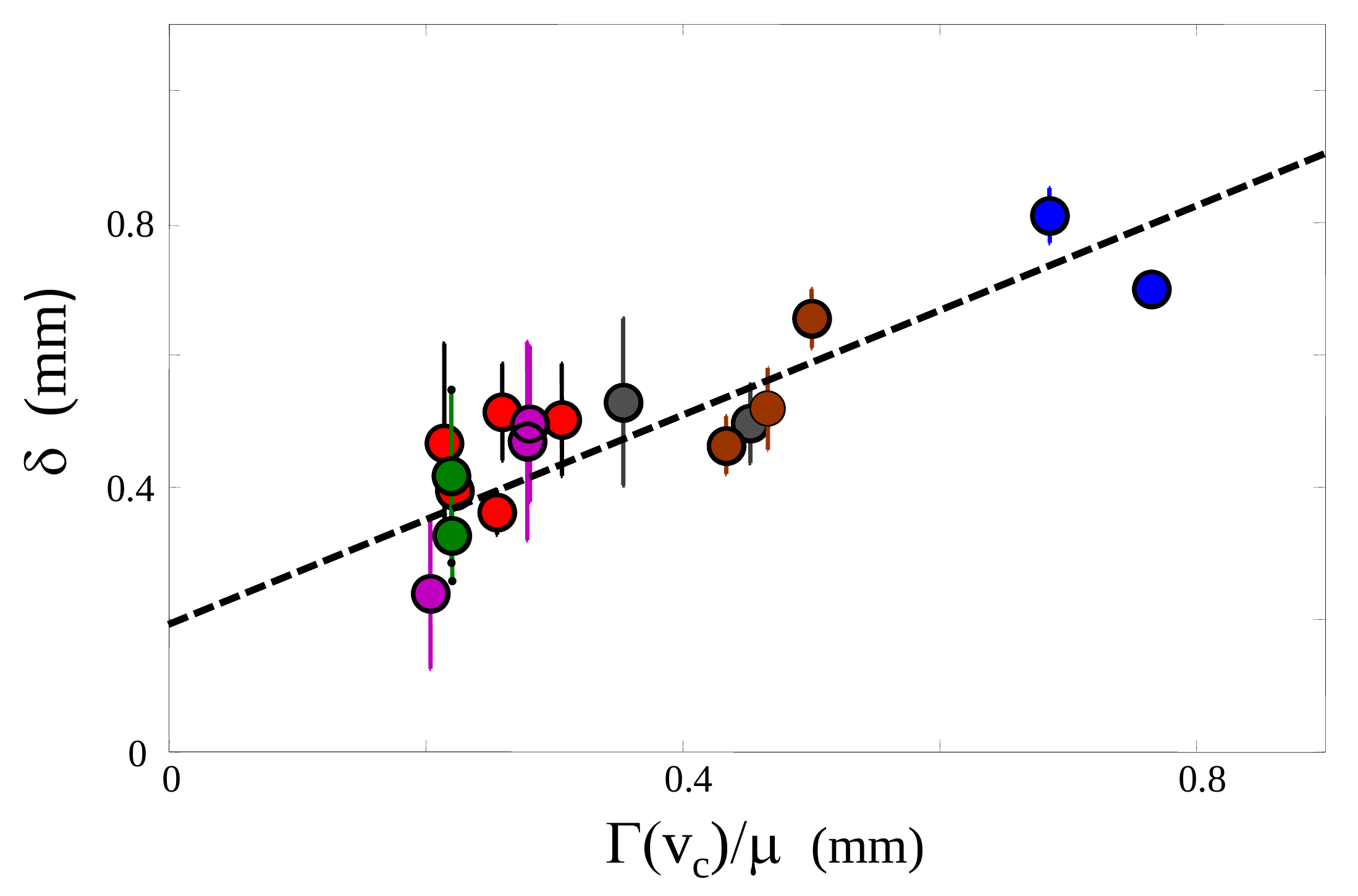} \caption{A comparison of the two estimates for the nonlinear scale, $\ell_{nl}$, for the different materials used (colors correspond to the legend in figure \ref{1st_wavelength}b). We see that both estimates, $\Gamma(v_c)/\mu$ and $\delta(v_c)$, are linearly dependent with approximately the same size. Note that the dashed line, which is a guide to the eye, indicates an {\em offset} value of $\sim\! 100\!-\!200\mu m$.}\label{nlscale}
\end{figure}
Comparing the two estimates of $\ell_{nl}$ in figure \ref{nlscale}, we see that they are both proportional to each other with nearly the same magnitude. The proportionality of these two scales, however, is marred by a slight offset of about $\sim 100-200\mu m$, as indicated by the dashed line in the figure. This is perhaps not surprising, since $\Gamma/\mu$ is an estimate of $\ell_{nl}$ predicted by the weakly non-linear theory, whereas $\delta(v_c)$ is a directly measured quantity that accounts for {\em all} nonlinear (i.e. deviations from linear elasticity) contributions to the crack tip scale (i.e. weak and strong elastic nonlinearities, dissipation etc.).

\begin{figure}
\centering \epsfig{width=.9\textwidth,file=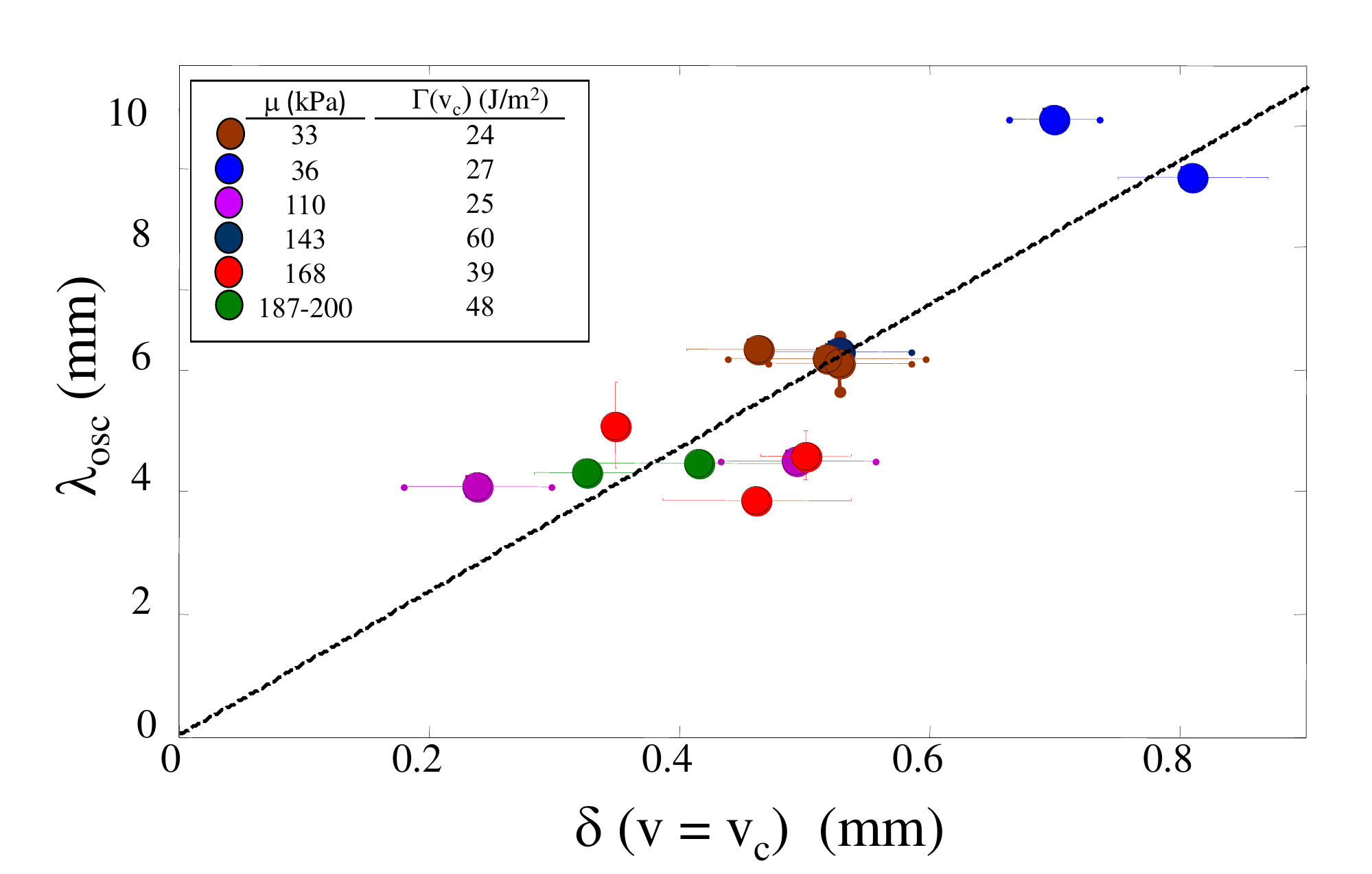} \caption{Comparison between the non-linear length-scale $\delta(v\!=\!v_c)$ and the oscillation wavelength $\lambda_{osc}$. Note that the different combinations of $\mu$, $\Gamma$, and $\epsilon$ are used to produce $\sim 15$ independent measurements. The dashed line is a guide to the eye. Data are taken from \cite{Goldman.12}.}\label{wavelength}
\end{figure}

In figure \ref{wavelength} we compare the wavelength of the first observed oscillation, $\lambda_{osc}$, which should correspond to the linearly unstable wavelength in the region of linear growth of the instability, to $\delta(v_c)$ for the 6 materials used. We indeed find that $\delta$ is directly proportional to $\lambda_{osc}$, as predicted in \cite{Bouchbinder.09b}.

In summary, the results of this section conclusively demonstrate that the oscillatory instability of fast brittle cracks indeed involves an intrinsic scale that is governed, in a large part, by the nonlinear elastic zone surrounding the crack tip. The size of this zone quantitatively agrees with the predictions of \cite{Bouchbinder.09b}. These results indicate that the nonlinear (and dissipative) zones surrounding the tip of a moving crack are not ``passive'' objects that are simply ``dragged along'' by the crack tip. Instead, as suggested by \cite{Gao.96, Buehler.03, Buehler2006, Livne.07, Bouchbinder.09b}, this region may play an active role in destabilizing crack motion. The demonstration of this, as presented in this work, is therefore an important step in obtaining a fundamental understanding of the origin of instabilities in dynamic fracture. These ideas are as general as the singular behavior that occurs at the tip of a moving crack. We believe that it is, therefore, quite likely that dynamics of the near-tip zone could play an important role in unraveling the physical mechanism driving other instabilities of rapid cracks \cite{Fineberg.91,Livne.07,Deegan.02.rubber,Ravi-Chandar.84b,Scheibert.2010}.

\section{Summary and open challenges}

In this article we first briefly reviewed the most well-developed theory of fracture -- Linear Elastic Fracture Mechanics (LEFM) -- and its major predictions. We argued that while this theory is very successful in various aspects, it falls short of explaining the fast dynamics of a crack once it deviates from a perfectly straight path. Thus, high-velocity path instabilities, most notably the side-branching and the oscillatory instabilities, remain open problems in this framework. We then summarized recent experimental and theoretical progress in understanding the dynamics of rapid brittle fracture, with a special focus on instabilities. We have highlighted the central role played by near crack front/tip nonlinearities and the associated intrinsic length scales in understanding these instabilities.

From an experimental perspective, we demonstrated that soft brittle elastomers mirror in detail the fracture phenomenology of more traditional brittle materials. By taking advantage of their significantly reduced wave-speeds we have been able to probe the brittle fracture process at length and time scales that were not  previously directly accessible. These experiments have led to two major outcomes. First, they enabled precise quantitative verification of detailed predictions of Linear Elastic Fracture Mechanics (LEFM) for the dynamics of straight cracks to an unprecedented degree and dynamical range. These experimental tests include both straight cracks propagating in an effectively infinite medium and straight cracks propagating in an infinite strip, where interactions with external boundaries qualitatively change the nature of the dynamics. Second, these experiments have revealed how linear elasticity breaks down near crack fronts/tips and clearly indicated the importance of near crack tip nonlinearities and the associated intrinsic length scales for crack instabilities, suggesting where essential physics is missing.

From a theoretical perspective, we described in detail the development of a new theoretical framework, the weakly nonlinear theory of fracture, which was directly motivated by the experimental observations on the breakdown of linear elasticity near crack fronts/tips. The basic premise of this theory is that the onset of this breakdown process is nonlinear elastic in nature. The predictions of this theory, in particular the form of the near crack front/tip singularity and the crack tip opening profile, are in excellent quantitative agreement with the experimental measurements.

This theory also predicts the existence of an intrinsic, i.e. geometry and loading independent, length scale that emerges from a competition between linear elastic and weakly nonlinear elastic deformations near crack fronts/tips. Based on the existence of this finite length scale, an equation of motion for the direction of crack propagation in 2D has been described. This extension of the nonlinear theory to path dynamics predicts a high-speed oscillatory instability whose wave-length is determined the intrinsic length scale. We have shown that this prediction is supported by experiments on a variety of different soft brittle materials. In our view, this is an important result that demonstrates that the near crack front/tip region not only accounts for the dissipation that accompany crack propagation, but may also play a central role in determining crack stability.

The work described in this paper is a basic attempt to account for elastic nonlinear response of materials in the vicinity of a crack tip. We have shown that new physical effects result when only {\em weak} elastic nonlinearity is taken into account. What happens when a material is so tough as to enable strong elastic nonlinearities to occur on large scales? One such example is the fracture of rubber, where strains larger than unity are easily obtained. When such enormous energy densities are reached, it is possible that an underlying premise of fracture mechanics, that energy must be transported from remote distances to enable fracture, should be re-examined. Work in this direction has, for example, demonstrated that in such circumstances supersonic tensile fracture can take place \cite{Marder.JMPS.2006,Marder.10, BFM10}.

The oscillatory instability discussed in detail in this review has been observed in a range of soft brittle materials, where cracks have been driven to extremely high velocities and in which elastic nonlinearities are both pronounced and spatially well-separated from dissipative nonlinearities. It remains to be seen whether it can be observed in more ``standard'' brittle materials, where no clear separation exists between nonlinear elastic scales and dissipative ones, in spite of the enormous technical difficulties involved. One may speculate that the existence of $\ell_{ln}$, whether associated with elastic or dissipative nonlinearities, is sufficient for the existence of the oscillatory instability. This challenge might be at least partially addressed through advanced numerical simulations.

Within a broader context, we believe that an additional crucial step in pushing the field forward is the substantiation and extension of the 2D equation of crack propagation and the development of the 3D counterpart for crack fronts. Addressing this challenge from both theoretical and experimental perspectives, and making progress along these lines, may open the way for more systematic path stability analysis and enhance our ability to quantitatively predict the failure dynamics of materials and interfaces. This may have important consequences for various problems in a wide range of fields, ranging from materials science to biology.

Such progress may also pave the way to resolve one of the most resistant puzzles in the field of dynamic fracture -- the side-branching instability. As was shown above, aspects of this instability possess an intrinsic 3D nature. One might then surmise that to ``crack" this instability one might have to further extend our understanding of 3D crack front dynamics \cite{Willis.02,Adda-Bedia.13,Henry.08}. The latter is intimately related to other open problems in the field, such as the roughness of crack surfaces \cite{bonamy.12, Bonamy.11,Ponson.10,Ponson.06}, crack front waves \cite{Morrissey2000,Sharon2002,Sherman.00,Willis.07} and the stability of crack fronts to an ensemble of continuous perturbations \cite{Roux.03, Ponson.13,Vandembroucq.04}.

\section*{Acknowledgments}

E. B. and J. F. acknowledge support from the James S. McDonnell Fund, E. B. acknowledges support from the Minerva Foundation with funding from the Federal German Ministry for Education and
Research, the Harold Perlman Family Foundation and the William Z. and Eda Bess Novick Young Scientist Fund. J. F. and T. G. acknowledge support from the European Research Council (Grant No. 267256), and the Israel Science Foundation (Grant 76/11).

\section*{References}

\bibliography{fracture_arxiv}
\bibliographystyle{unsrt}

\end{document}